%% file: ms.tex
\documentclass{emulateapj}
\usepackage{amssymb}
\usepackage{lscape}
\newcommand{\msun}{\ensuremath{\rm M_\odot}}
\newcommand{\msunyr}{\ensuremath{\rm M_{\odot}\;{\rm yr}^{-1}}}
\newcommand{\Ha}{\ensuremath{\rm H\alpha}}
\newcommand{\Hb}{\ensuremath{\rm H\beta}}

\newcommand{\Ntwo}{[\ion{N}{2}]}

\newcommand{\ztwo}{\ensuremath{z\sim2}}
\newcommand{\Othree}{[\ion{O}{3}]}
\newcommand{\Otwo}{[\ion{O}{2}]}

\begin{document}

\title{THE MASS-METALLICITY RELATION AT $z\gtrsim2$\altaffilmark{1}}
\author{\sc Dawn K. Erb\altaffilmark{2,3}, Alice
  E. Shapley\altaffilmark{4,5}, Max Pettini\altaffilmark{6},\\ Charles
  C. Steidel\altaffilmark{2}, Naveen A. Reddy\altaffilmark{2}, Kurt
  L. Adelberger\altaffilmark{7}} 

\shorttitle{THE MASS-METALLICITY RELATION AT $z\gtrsim2$}
\shortauthors{ERB ET AL.}

\slugcomment{Accepted for publication in \apj}

\altaffiltext{1}{Based on data obtained at the 
W.M. Keck Observatory, which is operated as a scientific partnership
among the California Institute of Technology, the University of
California, and NASA, and was made possible by the generous financial
support of the W.M. Keck Foundation.}  
\altaffiltext{2}{California Institute of Technology, MS 105--24,
  Pasadena, CA 91125} 
\altaffiltext{3}{Harvard-Smithsonian Center for Astrophysics, MS 20,
  60 Garden St, Cambridge, MA 02138; derb@cfa.harvard.edu} 
\altaffiltext{4}{Department of Astronomy, 601 Campbell Hall, University of
  California at Berkeley, Berkeley, CA 94720}
\altaffiltext{5}{Department of Astrophysical Sciences, Princeton
  University, Peyton Hall, Ivy Lane, Princeton, NJ 08544}
\altaffiltext{6}{Institute of Astronomy, Madingley Road, Cambridge CB3
  0HA, UK}
\altaffiltext{7}{Carnegie Observatories, 813 Santa Barbara Street,
  Pasadena, CA 91101}  

\begin{abstract}
We use a sample of 87 rest-frame ultraviolet-selected star-forming
galaxies with mean spectroscopic redshift $\langle z \rangle = 2.26
\pm 0.17$ to study the correlation between metallicity and stellar
mass at high redshift.  Using stellar masses determined from spectral
energy distribution fitting to $U_nG{\cal R}JK_s$ (and Spitzer IRAC,
for 37\% of the sample) photometry, we divide the sample into six bins
in stellar mass, and construct six composite \Ha~+~\Ntwo\ spectra from
all of the objects in each bin.  We estimate the mean oxygen abundance
in each bin from the \Ntwo/\Ha\ ratio, and find a monotonic increase
in metallicity with increasing stellar mass, from $12+\rm
log(O/H)<8.2$ for galaxies with $\langle M_{\star} \rangle = 2.7
\times 10^{9}$ \msun\ to $12+\rm log(O/H)=8.6$ for galaxies with
$\langle M_{\star} \rangle = 1.0 \times 10^{11}$ \msun.  The
mass-metallicity relation at \ztwo\ is offset from the local
mass-metallicity relation by $\sim0.3$ dex, in the sense that galaxies
of a given stellar mass have lower metallicity at high redshift.  A
corresponding metallicity-luminosity relation constructed by binning
the galaxies according to rest-frame $B$ magnitude shows no
significant correlation.  This lack of correlation is explained by the
known large variation in the rest-frame optical mass-to-light ratio at
\ztwo, and indicates that the correlation with stellar mass is more
fundamental.  We use the empirical relation between star formation
rate density and gas density to estimate the gas fractions of the
galaxies, finding an increase in gas fraction with decreasing stellar
mass.  The median gas fraction is more than two times higher than that
found in local star-forming galaxies, providing a natural explanation
for the lower metallicities of the \ztwo\ galaxies.  These gas
fractions combined with the observed metallicities allow the
estimation of the effective yield $y_{\rm eff}$ as a function of
stellar mass; in contrast to observations in the local universe which
show a decrease in $y_{\rm eff}$ with decreasing baryonic mass, we
find a slight increase.  Such a variation of metallicity with gas
fraction is best fit by a model with supersolar yield and an outflow
rate $\sim4$ times higher than the star formation rate.  We conclude
that the mass-metallicity relation at high redshift is driven by the
increase in metallicity as the gas fraction decreases through star
formation, and is likely modulated by metal loss from strong outflows
in galaxies of all masses.  Our ability to detect differential metal
loss as a function of mass is limited by the small range of baryonic
masses spanned by the galaxies in the sample, but there is no evidence
for preferential loss of metals from low mass galaxies as has been
suggested in the local universe.

\end{abstract}


\section{Introduction}
Correlations between mass and metallicity or luminosity and
metallicity are well-established in nearby galaxies, ranging over
orders of magnitude in mass and luminosity and spanning $\sim2$ dex in
chemical abundance.  \citet{lpr+79} first observed a correlation
between heavy element abundance and the total mass of galaxies; since
then, most investigations have focused on the metallicity-luminosity
relationship (e.g.\ \citealt{skh89,zkh94,gss97,lmc+04,slm+05}, to name
only a few), though others have studied correlations between
metallicity and rotational velocity \citep{zkh94,g02}.  The
relationship between metallicity and stellar mass has recently been
quantified by \citet[][T04 hereafter]{thk+04}, using a sample of
$\sim53,000$ galaxies from the Sloan Digital Sky Survey (SDSS).  Such
correlations provide insight into the process of galaxy evolution, as
they allow the study of the history of star formation and gas
enrichment or depletion through current, observable properties.
Chemical enrichment is a record of star formation history, modulated
by inflows and outflows of gas, while the stellar mass provides a more
straightforward measure of the accumulated conversion of gas into
stars and thus of the metals returned to the gas as the byproducts of
star formation.

It has long been recognized that a correlation between stellar mass
and gas phase metallicity is a natural consequence of the conversion
of gas into stars in a closed system \citep{v62,s63,ss72}.  The
principal ingredients of this ``simple'' or
``closed-box'' model are the metallicity, the yield from star formation
(defined as the mass of metals produced and ejected by star
formation, in units of the mass that remains locked in long-lived
stars and remnants), and the gas fraction.  If there are no 
inflows or outflows of gas, the metallicity is a simple function of
the yield and the gas fraction, and rises as the gas is converted into
stars and enriched by star formation according to the yield.  The
model is subject to the further assumptions that the system is
well-mixed at all times, that it begins as pure gas with primordial
abundances, that stellar evolution and nucleosynthesis take place
instantaneously compared to the timescale of galactic evolution (the
instantaneous recycling approximation), and that the IMF and the yield
in primary elements of stars of a given mass are constant.  This
simple model, in combination with the observation that lower mass
galaxies tend to have higher gas fractions \citep{md97,bd00}, results in a
correlation between metallicity and total mass.

One of the first, best-known failures of the simple model is the
so-called ``G-dwarf problem'': the closed-box model overpredicts the
number of low-metallicity stars observed in the solar neighborhood.
This is one aspect of the more fundamental problem that galaxies are
clearly not closed boxes.  On the one hand, infall and mergers are
essential aspects of galaxy formation (e.g. \citealt{pp75,no05}).  On
the other, galactic-scale winds driven by star formation (a process
generally referred to as ``feedback'' from star formation activity)
are a ubiquitous feature of starburst galaxies at low
(e.g.\ \citealt{ham90,lh96,m99,shc+04}, among many others) and high
\citep{pss+01,ssp+03,sci+03} redshifts.  Metals are detected in the
intergalactic medium (IGM; \citealt{essp00,ssr04}) at high redshifts,
and, at $z=2$--3, their locations are strongly correlated with the
distribution of observed star-forming galaxies \citep{assp03,ass+05}.
The potential of such supernova-powered winds to expel gas from
galaxies and thus modify their chemical evolution has been known for
some time; \citet{l74} showed how winds could account for the
mass-metallicity relation in elliptical galaxies by preferentially
ejecting metals from those with lower masses.  This provides an
alternative or additional explanation for the mass-metallicity
correlation, and its appeal has increased in recent years as the
physical evidence of feedback has multiplied and models of galaxy
formation have recognized its importance
(e.g.\ \citealt{hs03,bbf+03,dw03,nshm04,mqt05}).

Motivated by these competing or complementary theories of the origin
of the mass-metallicity relation, T04 use the statistical power of
$\sim53,000$ star-forming SDSS galaxies to revisit the correlation,
confirming its existence over $\sim3$ orders of magnitude in stellar
mass and 1 dex in metallicity.  Using estimates of the effective yield
as a function of baryonic mass, they see evidence for metal depletion
in low mass galaxies, finding that a galaxy with baryonic mass
$M\sim4\times10^9$ \msun\ loses half its metals, while low-mass dwarf
galaxies are five times more metal depleted than L$_{\ast}$ galaxies.
They also find that metal depletion occurs in galaxies with masses as
high as $10^{10}$ \msun, and interpret their results as a signature of
winds from an early starburst phase.

Given the strength of the mass-metallicity correlation, and its
plausible origin from either metal loss through winds or the change in
gas fraction as gas is converted to stars (in combination with higher
gas fractions in low mass galaxies), it is not unreasonable to suppose
that it may be present in galaxies at high redshift.  This has been
difficult to test, however.  The strong optical emission lines usually
used to determine metallicities shift into the infrared past $z\sim1$,
making the required large samples of spectra much more difficult to
acquire.  As luminosity can be determined with considerably greater
ease than stellar mass, the first efforts focused on the
luminosity-metallicity relation at high redshift
\citep{kk00,pss+01,sep+04}, finding that galaxies at $z>2$ are
overluminous for their metallicities when compared to local galaxies.
\citet{sep+04} also found that galaxies with $M_{\star}\sim10^{11}$
\msun\ have appproximately solar metallicity, while lower mass
galaxies have $Z\sim 0.5 Z_{\odot}$.

In this paper we study the relationships among stellar mass,
luminosity and metallicity at $z\gtrsim2$, using a sample of 87
star-forming galaxies with \Ha\ and \Ntwo\ spectra.  We describe our
sample selection, observations, and data reduction procedures in
\S\ref{sec:data}, and discuss our methods of determining stellar mass
and metallicity in \S\ref{sec:meas}.  In \S\ref{sec:massmet} we give
our results, and in \S\ref{sec:disc} we discuss their implications for
the origin of the mass-metallicity relation.  Our conclusions are
presented in \S\ref{sec:end}.  We use a cosmology with $H_0=70\;{\rm
  km}\;{\rm s}^{-1}\;{\rm Mpc}^{-1}$, $\Omega_m=0.3$, and
$\Omega_{\Lambda}=0.7$; in such a cosmology, the universe at $z=2.26$
(the mean redshift of our sample) is 2.9 Gyr old, or 21\% of its
present age.  For comparisons with solar metallicity, we use the most
recent values of the solar oxygen abundance, 12 +
log(O/H)$_{\odot}=8.66$, and the solar metal mass fraction
$Z_{\odot}=0.0126$ \citep{ags+04}.

\section{Sample Selection, Observations, and Data Reduction}
\label{sec:data}
The galaxies discusssed in this paper are drawn from a sample of 114
galaxies with \Ha\ spectra described in detail by
\citet{ess+06a,ess+06b}.  The galaxies were selected by their
rest-frame UV colors, and their redshifts were confirmed with
rest-frame UV spectra from the LRIS-B spectrograph on the 10-m Keck I
telescope; an overview of the \ztwo\ sample is given by
\citet{ssp+04}.  We then obtained near-IR spectra for a subset of
these galaxies.  Galaxies were chosen for near-IR spectroscopy for a
wide variety of reasons, the end result of which is a sample very
similar to the full set of galaxies with spectroscopic redshifts but
somewhat biased toward objects that are bright in $K$ or red in
${\cal}R-K$ or $J-K$.  A detailed discussion of the sample selection
and its relation to the larger sample of UV-selected galaxies is
provided by \citet{ess+06a}.  For measurements of metallicity in the
present paper, we include all objects with \Ha\ spectra and $K$-band
magnitudes (most have $J$ magnitudes as well, and 32 have also been
observed at 3.6, 4.5, 5.4 and 8.0 \micron\ with the IRAC camera on the
Spitzer Space Telescope), except those with AGN signatures in either
their rest-frame UV or optical spectra.

The \Ha\ spectra were obtained between May 2002 and September 2004
with the near-IR spectrograph NIRSPEC \citep{mbb+98} on the Keck II
telescope.  For the redshifts of the galaxies presented here,
\Ha\ falls in the $K$-band; most observations were conducted with the
N6 filter, which spans the wavelength range 1.558--2.315 \micron, and
in low-dispersion mode, which provides a resolution of $R\sim1400$.
The data were reduced using standard procedures described by
\citet{ess+03,ess+06a}, and flux-calibrated with reference to near-IR
standard stars.

The near-IR imaging was carried out with the Wide-field IR Camera
(WIRC, \citealt{wirc}) on the Palomar 5-m Hale telescope.  We obtained
$\sim9\arcmin\times9$\arcmin\ images to $K_s\sim22.5$ and $J\sim24$
in four fields, with a typical integration time of $\sim11$ hours in
each band per field.  Data reduction and photometry were performed as
described by \citet{ess+06a} and \citet{sse+05}.  For a description of
the  mid-IR IRAC data, reductions, and photometry, see \citet{bhf+04}, 
\citet{sse+05}, and \citet{res+05}.

\section{Measurements}
\label{sec:meas}
\subsection{Stellar Masses}
\label{sec:massmeas}
Stellar masses are determined by fitting model SEDs to the $U_nG{\cal
  R}JK$ (and IRAC, when present) photometry, using the procedure
described in detail by \citet{sse+05} and \citet{ess+06a}, which uses
the \citet{bc03} population synthesis models and the \citet{cab+00}
extinction law.  We compare the model SEDs of galaxies with a variety
of ages and amounts of extinction to our observed photometry, and
obtain the star formation rate and stellar mass from the normalization
of the best-fit model to the data.  We try models with a constant star
formation (CSF) rate and models in which the star formation rate
smoothly declines with time, parameterized by ${\rm SFR}\propto
e^{-t/\tau}$, with $\tau=$10, 20, 100, 200, 500, 1000, 2000 and 5000
Myr.  In practice, however, most star formation histories provide
adequate fits to most objects, and we therefore use the CSF models
unless one of the $\tau$ models provides a significantly better fit.
The best-fit models for the galaxies discussed here are given by
\citet{ess+06a}.  We use a \citet{c03} IMF for the stellar masses and
star formation rates, which results in stellar masses and SFRs 1.8
times smaller than those computed using a \citet{s55} IMF.  These
stellar masses are the integral of the star formation rate over the
lifetime of the galaxy, and so represent the total mass of stars
formed rather than the mass in living stars at the time of
observation.  For our adopted IMF, the current living stellar mass is
$\sim10$--40\% lower depending on the age of the galaxy.  We use the
total rather than current stellar mass because the relevant quantity
for the simple chemical evolution models we apply in \S\ref{sec:disc}
is the fraction of the initial gas mass that has been turned into
stars.

Uncertainties in the fitting are determined from a large number of
Monte Carlo simulations in which the input photometry is varied
according to the photometric errors.  The simulations also take into
account variations in the star formation history by treating $\tau$ as
a free parameter with the possible values given above.  As has often
been noted for SED modeling of this sort
(e.g.\ \citealt{pdf01,ssa+01,sse+05}), the stellar mass is the most
securely-determined parameter.  For the current sample, the mean
fractional uncertainty $\langle \sigma_{M_{\star}}/M_{\star} \rangle =
0.4$.  The simulations and their results are described in more detail
by \citet{ess+06a}.

As discussed by \citet{pdf01} and \citet{sse+05}, one limitation of
such modeling is the insensitivity of the data to faint, old stellar
populations, which could be obscured by current star formation and
thus lead to an underestimate of the stellar mass.  We have estimated
the magnitude of this effect by fitting a variety of two-component
models to the observed SEDs, in which the light from the
observed-frame $K$-band and redward is constrained to come from a
maximally old burst while a young population is fitted to the
rest-frame UV residuals.  Such models make little difference to the
stellar masses of already massive galaxies, since their stellar
populations already approach the maximum age allowed by the age of the
universe at their redshift.  The masses of low-mass galaxies can be
increased by an order of magnitude by such models, but the models are
generally a poor fit to the SED and result in star formation rates far
higher than those determined by all other indicators, with an average
SFR of $\sim900$ \msunyr.  While we cannot rule out such models in
individual cases, they are very unlikely to be correct on average, and
we therefore consider it unlikely that we have underestimated the
stellar masses of the low-mass galaxies by such a large factor.  More
general two-component models, in which the relative contributions from
a maximally old population and a young burst are allowed to vary,
increase the stellar mass by a factor of a few at most.  The
two-component models are described in more detail by \citet{ess+06a}.

As described below, for the purposes of determining metallicities we
divide the sample into six bins by stellar mass, with 14 or 15
galaxies in each bin.  The mean stellar mass in each bin ranges from
$2.7\times10^9$ \msun\ to $1.1\times10^{11}$ \msun.  The means and
standard deviations of the fitted parameters for each bin are given in
Table~\ref{tab:props}, along with the star formation rates determined
from \Ha\ luminosities.  The \Ha\ luminosities have been corrected for
extinction using the \citet{cab+00} extinction law and the best-fit
value of $E(B-V)$ from the SED modeling, and we have applied a factor
of two aperture correction determined from the comparison of the
NIRSPEC spectra and narrowband images (see \citealt{ess+06b} for
details, and for a full discussion of the \Ha-derived star formation
rates and their implications).

\subsection{Metallicities}
\label{sec:metalmeas}
The most direct way to determine the abundances of metals from the
observed emission line fluxes in \ion{H}{2} regions is through the
measurement of the electron temperature $T_e$.  As the metallicity of
the gas increases, the cooling through metal emission lines also
increases, resulting in a decrease in $T_e$.  The ratio of the auroral
(the transition from the second lowest to the lowest excited level)
and nebular (the transition from the lowest excited level to the
ground state) emission lines of the same ion is highly sensitive to
the electron temperature, and therefore the measurement of such pairs
of lines has been the preferred method of determining abundances in
\ion{H}{2} regions.  However, the auroral lines (in particular the
most widely used line, \Othree$\lambda4363$) become extremely weak at
metallicities above $\sim0.5$ solar, and undetectable at all
metallicities in the low S/N spectra of distant galaxies.  In most
cases, therefore, we must use the empirical ``strong line'' abundance
indicators, which are based on the ratios of collisionally excited
forbidden lines to hydrogen recombination lines.

These are calibrated with reference to the $T_e$ method or, more
commonly, with detailed photoionization models.  The strong line
methods carry significant hazards, however.  Substantial biases and
offsets are observed between abundances determined with different
methods, and between different calibrations of the same method
(e.g.\ \citealt{kbg03,kk04}), and many of the strong line indicators
are sensitive to the ionization parameter as well as metallicity.  The
advent of large telescopes, sensitive detectors, and spectrographs
with high throughput has enabled the measurement of abundances with
the $T_e$ method in an increasingly large sample of extragalactic
\ion{H}{2} regions.  These new data suggest that the most widely used
indicator, $R_{23}\equiv
($\Otwo$\lambda3727+$\Othree$\lambda\lambda4959,5007)/\Hb$, may
systematically overestimate metallicities in the high-abundance regime
($Z\gtrsim Z_{\odot}$) by as much as 0.2--0.5 dex
\citep{kbg03,gkb04,bsgs05}.  The $T_e$ method itself is not without
difficulties, however, as temperature gradients or fluctuations in the
\ion{H}{2} regions may affect abundance determinations at high
metallicities \citep{s05}.  The result of all this is that absolute
values of metal abundances are still quite uncertain; but, fortunately
for our present purposes, relative abundances of similar objects,
determined with the same method, are more reliable.
 
The available data limit the options for determining the chemical abundances of the
\ztwo\ galaxies.  NIRSPEC can observe only one
band in a single exposure, and at these redshifts \Ha\ and
\Ntwo$\lambda6584$ fall in the $K$-band, [\ion{O}{3}]$\lambda\lambda$5007,4959 and
\Hb\ in the $H$-band, and [\ion{O}{2}]$\lambda$3727 in $J$.  We have
focused our observations on \Ha\ in the $K$-band, so obtaining the
additional data needed to use the $R_{23}$ indicator
would triple the required observing time.  Our only option to
determine the metallicity of the vast majority of the galaxies in our
sample is thus the ratio of \Ntwo\ to \Ha.  In addition to minimizing
the observing time required to obtain a large sample, this ratio has
the advantages that it is insensitive to reddening and is not affected by
the relative uncertainties in flux calibration of spectra taken in
different bandpasses.

The use of $N2\equiv$ log \Ntwo$\lambda6584$/\Ha\ as an abundance
indicator was proposed by \citet{sck94}, and has been further
discussed and refined by \citet{rsb+00}, \citet{dtt02} and
\citet{pp04}.  The \Ntwo/\Ha\ ratio is affected by metallicity in two
ways.  First, there is a tendency for the ionization parameter of an
\ion{H}{2} region to decrease with increasing metallicity (see, for
example, Figure 3 of \citealt{d05}), thereby increasing the ratio
\Ntwo/[\ion{N}{3}] and hence \Ntwo/\Ha.  Second, nitrogen has both a
primary component whose abundance varies at the same rate as other
primary elements, and a secondary component which increases in
abundance with increasing metallicity, further raising the
\Ntwo/\Ha\ ratio.  Using electron temperature measurements to
determine ionic abundances in 20 extragalactic \ion{H}{2} regions,
\citet{kbg03} find that the nitrogen abundance is adequately described
by a simple model with a primary component with constant $\rm
log(N/O)=-1.5$ and a secondary component for which $\rm
log(N/O)=log(O/H)+2.2$.  In other words, secondary nitrogen becomes
important for $\rm 12+log(O/H)\gtrsim 8.3$; unless we have
significantly overestimated the metallicities of the galaxies in our
sample, most of our objects are in this regime.  As emphasized by
\citet{kd02}, a drawback of the $N2$ indicator is its sensitivity to
the ionization parameter as well as to metallicity.  In addition, the
\Ntwo/\Ha\ ratio cannot be used to determine metallicities above
approximately solar, at which the $N2$ index saturates as nitrogen
becomes the dominant coolant.

\begin{figure}[htbp]
\plotone{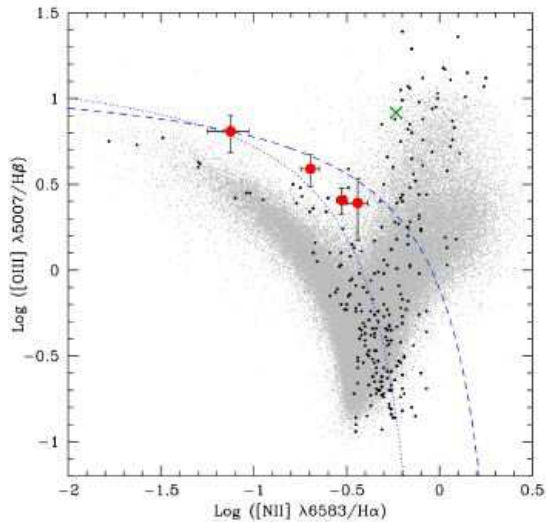}
\caption{The \Othree/\Hb\ vs.\ \Ntwo/\Ha\ diagnostic diagram.  The
  four galaxies in our sample for which we have measurements of all
  four lines are shown by the large red circles.  The $z=2.2$ galaxy
  discussed by \citet{vkr+05}, which shows evidence of an AGN or shock
  ionization by a wind, is shown by the green $\times$.  The small
  grey points represent $\sim96,000$ objects from the SDSS, and the
  small black points are the local starburst galaxies studied by
  \citet{khd+01}.  The dashed line shows the maximum theoretical
  starburst line of \citet{kds+01}; for realistic combinations of
  metallicity and ionization parameter, star-forming galaxies fall
  below and to the left of this line.  The dotted line is a similar,
  empirical determination by \citet{kht+03}.}
\label{fig:diag}
\end{figure}

Finally, the \Ntwo/\Ha\ ratio is sensitive to contamination from AGN
and shock excitation, and if possible these should be ruled out before
using it as a metallicity indicator.  This is generally done through a
diagnostic line ratio diagram such as
\Othree/\Hb\ vs.\ \Ntwo/\Ha\ \citep{bpt81,vo87}; on such a diagram,
normal galaxies and AGN fall in generally well-defined regions.
Unfortunately we lack the data to place nearly all of our objects on
such a diagram; because we have so far obtained very few
\Othree/\Hb\ spectra, we have measurements of all four lines for only
four galaxies.  The ratios \Othree/\Hb\ vs.\ \Ntwo/\Ha\ for these
galaxies are shown in Figure~\ref{fig:diag}, along with $\sim96,000$
objects from the SDSS (small grey points), the local starbursts
analyzed by \citet{khd+01} (small black points), and the $z=2.2$
galaxy (green \boldmath$\times$\unboldmath) discussed by
\citet{vkr+05}, which shows evidence of shock ionization or an AGN and
which falls in a clearly different region of the diagram than our
galaxies.  The dashed blue line shows the maximum theoretical
starburst line determined from photoionization modeling by
\citet{kds+01}; for realistic combinations of metallicity and
ionization parameter, models of normal starbursts fall below and to
the left of this line.  The lower blue dotted line is a similar,
empirically determined classification line derived for the SDSS
objects by \citet{kht+03}.

Our galaxies fall between the two lines, along a sequence with a
higher \Othree/\Hb\ ratio at a given \Ntwo/\Ha\ ratio (or a higher
\Ntwo/\Ha\ ratio for a given \Othree/\Hb\ ratio) compared to the SDSS
galaxies\footnote{We have not corrected the \Hb\ fluxes for stellar
  absorption, but we do not expect this to be a significant effect.
  We expect the stellar \Hb\ absorption line to have an equivalent
  width $W_{\rm abs}\gtrsim -5$ \AA\ \citep{khd+01,kkp99}, while,
  assuming a typical ratio of $W_{\Ha}/W_{\Hb}\sim5$ \citep{sb99}, our
  galaxies have $W_{\Hb}\sim 50$ \AA.}.  A similar offset is seen in
star-forming galaxies at $z\sim1.4$ \citep{scmb05}.  The SDSS contains
very few objects in this region, indicating that there are important
physical differences between the local and high redshift samples.  One
plausible way to produce such a shift in the
\Ntwo/\Ha--\Othree/\Hb\ diagram is through some combination of a
harder ionizing spectrum and an increase in electron density.  Though
they are weak, the density-sensitive [\ion{S}{2}] lines in the
composite spectra indicate an average electron density of $n_e\sim500$
cm$^{-3}$ (with no dependence on stellar mass), higher than that found
in normal local galaxies but similar to densities seen in local
starbursts \citep{khd+01}. Furthermore, \citet{kds+01} show that a
relatively hard EUV radiation field is required to model the line
ratios of local starbursts; note that a shift is also observed between
the local starbursts and the SDSS sample.  Detailed photoionization
modeling is required to determine quantitatively the origin of this
shift and its effect on the calibration of the $N2$ index, and for the
moment the absolute calibration of the metallicity scale remains
uncertain (though the relatively shallow slope of the relation between
$N2$ and (O/H) means that offsets are unlikely to be large).  As
diagnostic line ratios for more high redshift galaxies are measured,
it is hoped that full photoionization modeling, with improved spectra
of the Wolf-Rayet stars that dominate the EUV radiation field, will
clarify the physical reasons at the root of the shifts in the
\Ntwo/\Ha--\Othree/\Hb\ diagram.

We can exclude an AGN contribution to our sample as a whole based on
the X-ray and UV properties of the galaxies. Specifically,
\citet{res+05} find that only 3\% of UV-selected galaxies at $z \sim
2$ have X-ray detections in the ultra-deep 2\,Ms {\em Chandra} Deep
Field North. These galaxies, almost all of which have $K_s < 20$, are
not included in the composite spectra considered in the present
analysis.  Further information on AGN contamination comes from the
rest-frame UV spectra of our galaxies, which allow the rejection of
AGN on an individual basis in fields without deep X-ray data.  As
described further in \S\ref{sec:massmet}, we have constructed six
composite UV spectra, binned by stellar mass in the same way as the
NIRSPEC spectra (see below), as well as larger composites of the two
highest and two lowest mass bins.  None of the composites show AGN
features such as broad and/or high ionization emission lines, placing
further limits on low-level AGN activity.  On the basis of all of
these considerations, we conclude that a significant AGN component to
the galaxies considered here is unlikely, and proceed under the
assumption that the emission lines we observe are produced in
\ion{H}{2} regions photoionized by hot stars.

This work is primarily concerned with relative, average abundances
determined from composite spectra (see below), which can be determined
with more accuracy than the metallicities of individual objects; this
mitigates the uncertainties associated with the $N2$ method.  We use
the calibration of \citet{pp04}, which is based on \ion{H}{2} regions
whose ionic abundances have been determined from the electron
temperature or from detailed photoionization modeling.  From a sample
of 137 such \ion{H}{2} regions, 131 of which have abundances
determined with the $T_e$ method, \citet{pp04} find
\begin{equation}
12+{\rm log (O/H)}=8.90+0.57\times N2,
\end{equation}
with a 1$\sigma$ dispersion of 0.18 dex.  They conclude that the $N2$
calibrator allows a determination of the oxygen abundance to within a
factor of $\sim2.5$ with 95\% confidence, an accuracy comparable to
that of the $R_{23}$ method. 

\begin{figure}[htbp]
\plotone{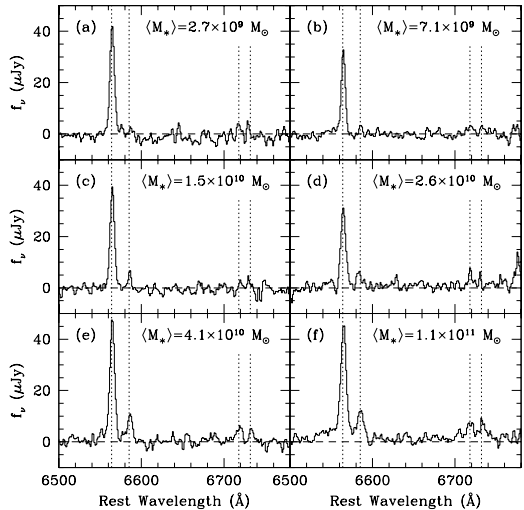}
\caption{The composite NIRSPEC spectra of the 87 galaxies in our
  sample, divided into six bins of 14 or 15 objects each by increasing
  stellar mass (panels {\it a} through {\it f}).  The spectra are
  labeled with the mean stellar mass in each bin, and the \Ha, \Ntwo,
  and [\ion{S}{2}] lines are marked with dotted lines (left to right
  respectively).  The increase in the strength of \Ntwo\ with stellar
  mass can be seen clearly.  The density-sensitive [\ion{S}{2}] lines,
  while weak, indicate a typical electron density of $n_e\sim500$
  cm$^{-3}$, with no significant dependence on mass; this is a value
  comparable to that seen in local starburst galaxies \citep{khd+01}.}
\label{fig:spectra}
\end{figure}

A further difficulty of the $N2$ method is that the \Ntwo\ line is
weak, and is generally detected in the spectra of individual objects
in our sample only when they approach solar metallicity \citep{sep+04}
or have especially strong line fluxes.  In order to increase the S/N
and improve the likelihood of detecting \Ntwo\ at lower metallicities,
we have divided the sample into six bins by stellar mass, with 14 or
15 galaxies in each bin, and constructed a composite \Ha\ +
\Ntwo\ spectrum of the objects in each bin.  The use of composite
spectra has the advantage that, in addition to increasing S/N, it
minimizes the effects of the mass and metallicity uncertainties of
individual objects, because we are only concerned with the average
properties of the galaxies in each bin.  We first shift the
flux-calibrated spectra into the rest-frame, and then average them,
rejecting the minimum and maximum value at each dispersion point to
suppress noise from large sky subtraction residuals in individual
spectra.  The six composite spectra, labeled with the mean stellar
mass in each bin, are shown in Figure~\ref{fig:spectra}.

We measure the \Ntwo/\Ha\ ratio by first measuring the \Ha\ fluxes,
central wavelengths, and widths, and then constraining the \Ntwo\ line
to have the same width and a central wavelength fixed by the position
of \Ha.  We use the rms of the spectrum between emission lines to
determine the typical noise in each spectrum; because the galaxies are
at different redshifts, the systematic effects of the night sky lines
are minimized in the composites, and the rms provides an adequate
description of the noise.  This procedure provides a good fit to the
\Ntwo\ line for all of the composite spectra (except in the lowest
mass bin, where we determine an upper limit on the \Ntwo\ flux).  The
measured \Ha\ and \Ntwo\ fluxes for each composite, and the inferred
value of 12+log(O/H), are given in Table~\ref{tab:metals}.  The listed
uncertainties in 12+log(O/H) include the scatter in the $N2$
calibration (a 1$\sigma$ uncertainty of 0.18 dex reduced by $\sqrt N$,
where $N$ is the number of objects in the composite spectrum) as well
as the uncertainties in the measurements of the \Ha\ and
\Ntwo\ fluxes.

\section{The Mass-Metallicity Relation}
\label{sec:massmet}

\begin{figure*}[htbp]
\plotone{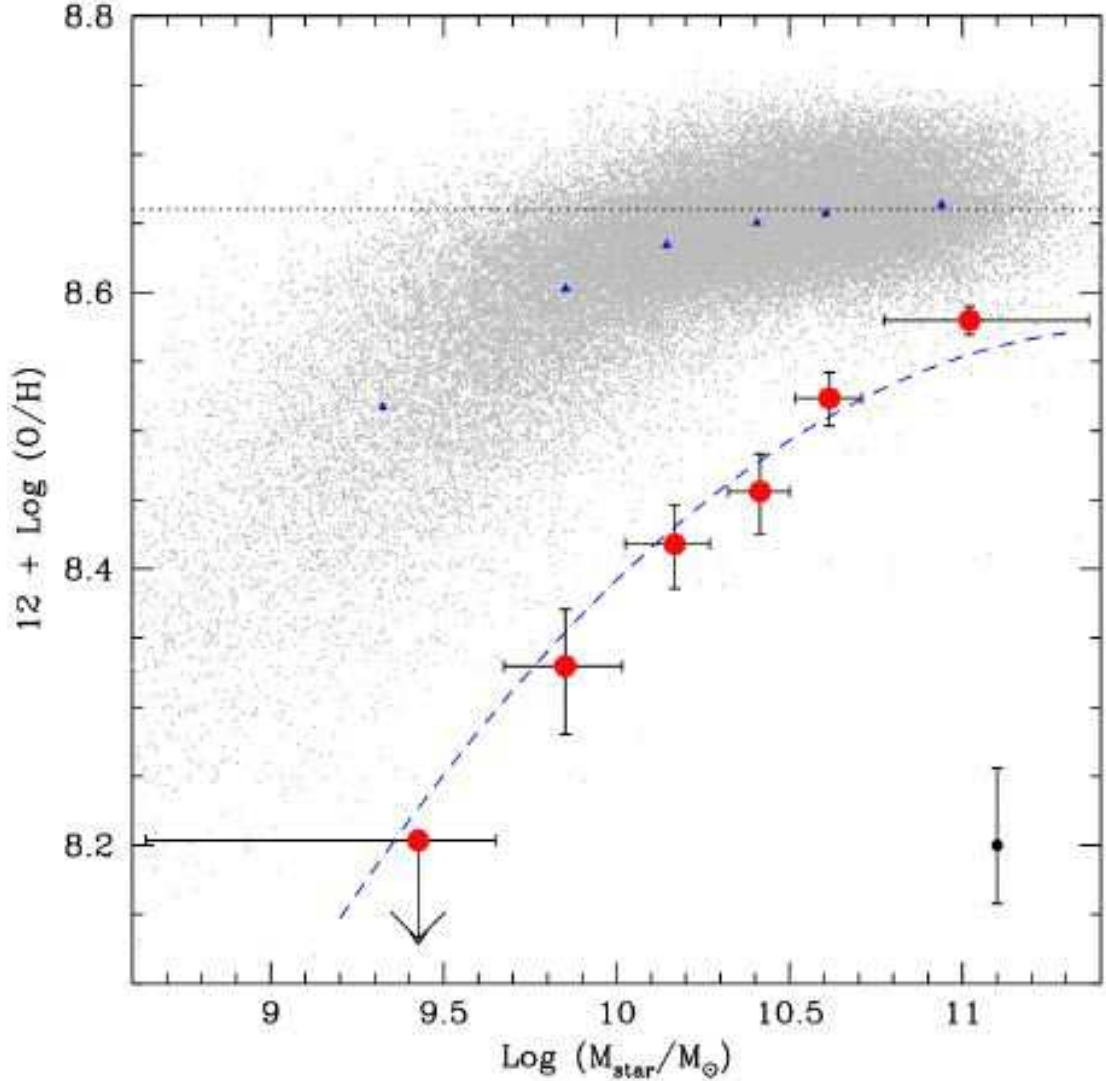}
\caption{The observed relation between stellar mass and oxygen
  abundance at $z\sim2$ is shown by the large red circles.  Each point
  represents the average value of 14 or 15 galaxies, with the
  metallicity estimated from the \Ntwo/\Ha\ ratio of their composite
  spectrum.  Horizontal bars indicate the range of stellar masses in
  each bin, while the vertical error bars show the uncertainty in the
  \Ntwo/\Ha\ ratio.  The additional error bar in the lower right
  corner shows the additional uncertainty in the $N2$ calibration
  itself.  The dashed blue line is the best-fit mass-metallicity
  relation of \citet{thk+04}, shifted downward by 0.56 dex.  The
  metallicities of different samples are best compared using the same
  calibration; we therefore show, with small grey points, the
  metallicities of the $\sim53,000$ SDSS galaxies of \citet{thk+04}
  determined with the $N2$ index.  Note that the \Ntwo/\Ha\ ratio
  saturates near solar metallicity (the horizontal dotted line).  The
  blue triangles indicate the mean metallicity of the SDSS galaxies in
  the same mass bins we use for our sample; using the more reliable,
  low metallicity bins, our galaxies are $\sim0.3$ dex lower in
  metallicity at a given mass.}
\label{fig:metalmass}
\end{figure*}

Figure~\ref{fig:metalmass} shows the mean metallicity of the galaxies
in each mass bin plotted against their mean stellar mass (large filled
circles); there is a monotonic increase in metallicity with stellar
mass.  The vertical error bars show the uncertainty in $\rm
12+log(O/H)$ from measurement uncertainties in the \Ntwo/\Ha\ ratio,
while the additional vertical error bar in the lower right corner
shows the uncertainty due to the scatter in the $N2$ calibration.  The
horizontal bars show the range of stellar masses in each bin.  The
most massive galaxies in the sample have close to solar metallicities,
a result found previously by \citet{sep+04}, who measured the
\Ntwo/\Ha\ ratio from individual spectra of the brightest objects with
$K_s<20$.  More typical galaxies with $M_{\star}\sim 10^{10}$
\msun\ have $\rm 12+log(O/H)\sim8.4$, while for the lowest mass
objects we can only place an upper limit $\rm 12+log(O/H)<8.2$, or
$\rm (O/H)<1/3\; (O/H)_{\odot}$.

We have not found any systematic effects which could spuriously
produce the clear observed correlation between stellar mass and
metallicity.  Specifically, it is possible that the value of
$M_{\star}$ in the lowest mass bins has been underestimated if an
older stellar population is already in place in these galaxies (see
the discussion in \S\ref{sec:massmeas}).  This would have the effect
of steepening the observed correlation, as the lowest mass bins in
Figure~\ref{fig:metalmass} would move to the right by 0.3--0.5 dex,
while the high mass bins would remain unaffected.  The use of the
integrated star formation rate as the stellar mass rather than
the mass in currently existing stars (see \S\ref{sec:massmeas})
would also steepen and shift the correlation somewhat, with the lower
mass points moving $\sim0.1$ dex to the left and the upper mass points
shifting left by $\sim0.2$ dex.  AGN contamination is highly unlikely
to produce the correlation, given the low fraction of AGN in the
sample.  Variations in the ionization parameter are also unlikely to
be correlated with the assembled stellar mass.  The ionization
parameter of an \ion{H}{2} region depends on the age of the ionizing
cluster, which is very much less than the age of the galaxy (see,
e.g., \citealt{d05}); since each galaxy in our sample presumably
contains many \ion{H}{2} regions of different ages, the overall
variation in ionization parameter from galaxy to galaxy should be
small.  There is also no dependence of electron density on galaxy
mass; the density-sensitive [\ion{S}{2}] lines in the composite
spectra indicate an average density $n_e\sim500$ cm$^{-3}$, with no
trend with stellar mass.  We therefore have no reason to expect the
ionization parameter to depend on the total stellar mass (other than
the dependence on metallicity, which is included in the $N2$
calibration).

The dashed line in Figure~\ref{fig:metalmass} shows the
mass-metallicity relation determined for $\sim53,000$ star-forming
SDSS galaxies by T04, after an arbitrary downward shift of 0.56 dex.
With this shift the SDSS relation matches the
\ztwo\ galaxies remarkably well, though it is slightly shallower.
The empirical shift of 0.56 dex includes an offset due to the
different abundance diagnostics used in the two studies---while the
SDSS metallicity determinations take into consideration all of the
strong nebular lines, ours are based on the $N2$ index alone, for the
reasons explained above.  For a more consistent comparison,
we use the $N2$ calibration to calculate the 
metallicities of the same 53,000 SDSS galaxies, shown with small
grey points in Figure~\ref{fig:metalmass}; the mean $N2$
metallicity, in bins spanning the same range of stellar masses used
for our sample, is shown by the small blue triangles.  The saturation
of the \Ntwo/\Ha\ ratio at metallicities approaching solar (the horizontal
dotted line) is clearly apparent in the SDSS sample, and makes the
determination of the true offset between the two samples 
difficult.  For galaxies with $M_{\star}\sim2\times10^9$ \msun, the
oxygen abundance of the \ztwo\ sample is lower by $>0.3$ dex; at
higher stellar masses the offset is smaller, due at least in part to
the saturation of the \Ntwo/\Ha\ ratio.  Taking the two lowest
mass bins as the most reliable indicators and assuming that the shape
of the relation remains unchanged, we find that
galaxies at \ztwo\ are $\sim0.3$ dex lower in metallicity than
galaxies of the same stellar mass today.  We discuss the likely
reasons for this offset, and its uncertainties, in \S\ref{sec:evol}. 

\subsection{Composite Ultraviolet Spectra}

Given the uncertainties in the absolute metallicity scale associated
with the $N2$ index, it would be highly desirable to obtain
independent abundance measures for the galaxies considered here.  As
discussed by \citet{rpl+04}, the ultraviolet spectrum of star-forming
galaxies is rich in stellar spectral features which provide abundance
diagnostics for the young stellar populations.  The difficulty is that
these are low-contrast features, usually requiring data of higher
quality than can be obtained with current instrumentation.
Nevertheless, it is worthwhile to examine whether the rest-frame
ultraviolet spectra of the galaxies under study are consistent with
the abundance trend revealed by Figure~\ref{fig:metalmass}.

\begin{figure*}[htbp]
\includegraphics[angle=-90,width=\textwidth]{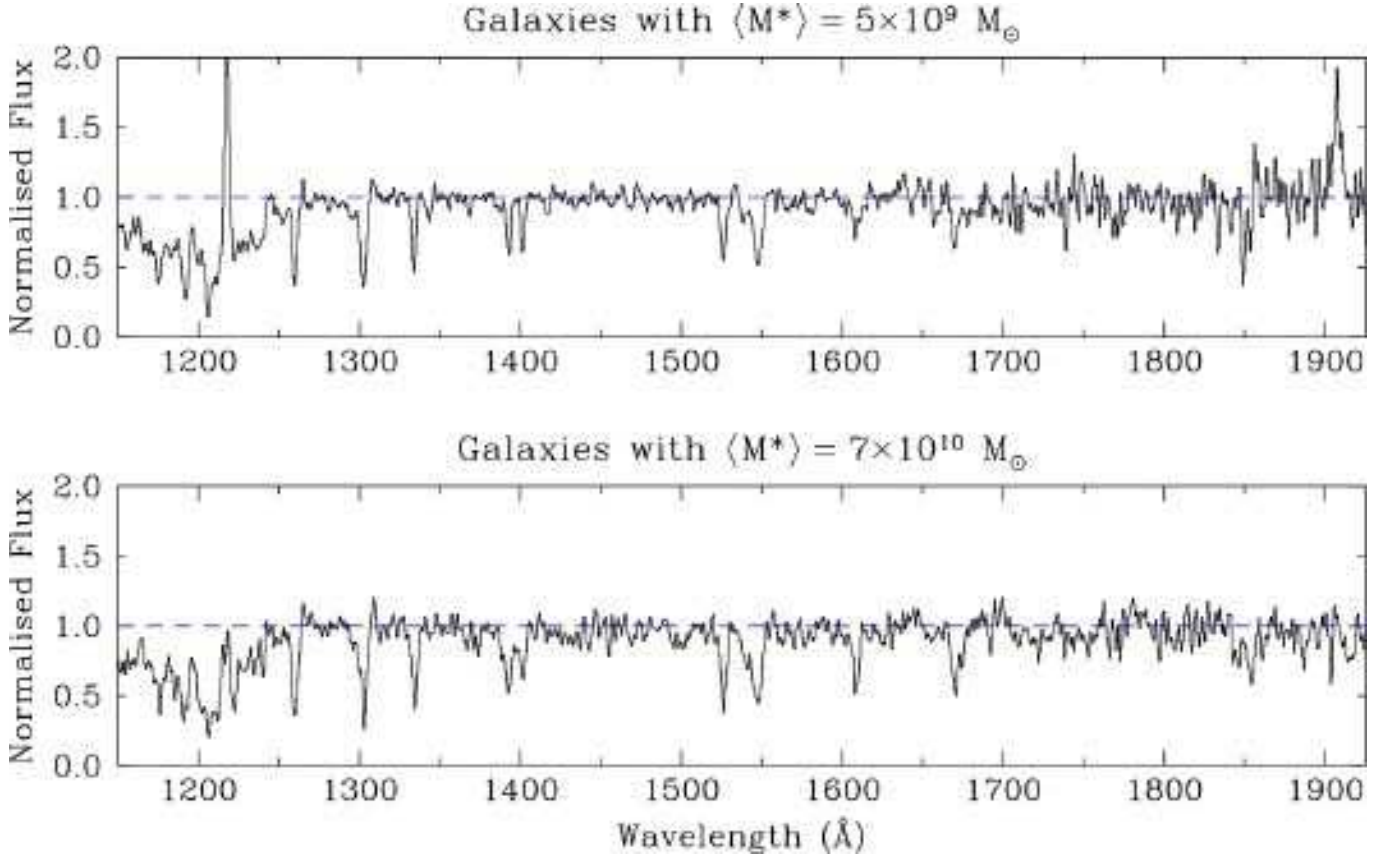}
\caption{Composite rest-frame UV spectra of galaxies in the
two lower (top panel) and two higher (bottom panel) mass
bins in our sample. The upper spectrum is the average of 30 LRIS-B
spectra of galaxies with mean stellar mass $\langle M_{\star} \rangle = 5
\times 10^{9}$ \msun,
while 28 spectra contribute to the lower composite
for which $\langle M_{\star} \rangle = 7 \times 10^{10}$ \msun.
The spectra have been divided by the underlying stellar continuum
estimated according to the prescription by \citet{rpl+04}.
Differences in the stellar, interstellar, and nebular lines between
the two composites are discussed in the text.}
\label{fig:uvspec}
\end{figure*}

To this end we have constructed two composite spectra, each consisting
of approximately 30 galaxies, by averaging the LRIS-B spectra of the
galaxies in, respectively, the two lower and the two higher mass bins
in Figure 3. The corresponding mean stellar masses are $\langle
M_{\star} \rangle = 5 \times 10^9$ \msun\ and $\langle M_{\star}
\rangle = 7 \times 10^{10}$ \msun.  The coarser mass binning was
required to improve the S/N of the rest-UV composites to the level
where the stellar absorption features which are sensitive to
metallicity could be clearly discerned.  The two composite spectra are
shown in Figure~\ref{fig:uvspec}, after normalization to the
underlying stellar continua following the prescription by
\citet{rpl+04}.

\begin{figure}[htbp]
\plotone{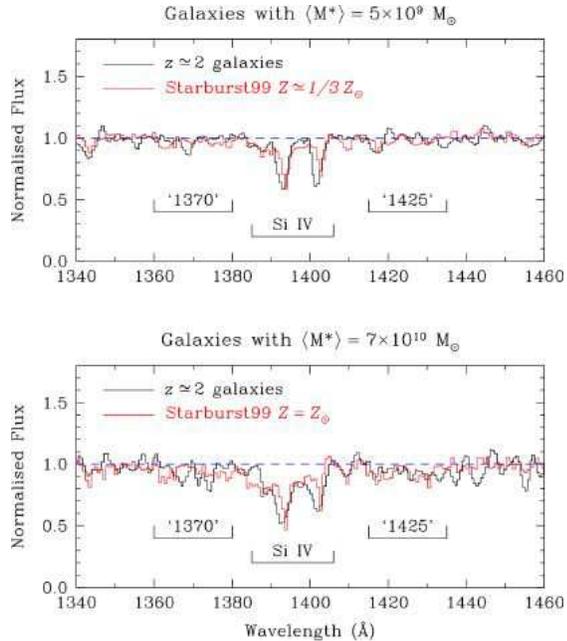}
\caption{Close-up of the 1400\,\AA\ region in the two composite
spectra shown in Figure~\ref{fig:uvspec} (black histogram).
This region contains two blends of stellar
photospheric lines, labeled ``1370'' and ``1425'', whose strength
is thought to depend primarily on metallicity, and
the \ion{Si}{4}~$\lambda\lambda 1393,1402$ doublet
which consists of a broad P-Cygni stellar absorption on which
narrower interstellar lines are superposed.
The red histogram shows the Starburst99 spectrum for the standard
case of continuous star formation with a Salpeter IMF and
 `Magellanic Cloud' (upper panel) or solar (lower panel) metallicities.}
\label{fig:uvzoom}
\end{figure}

Within the spectral region of 1150--1925\,\AA\ covered by the
composites, the interval near 1400\,\AA\ is particularly suitable for
an abundance analysis; we show this portion on an expanded scale in
Figure~\ref{fig:uvzoom}.  The region includes two blends of stellar
photospheric lines, centered at 1370\,\AA\ and 1425\,\AA, whose
strengths were shown by \citet{llh+01} to be mostly sensitive to
metallicity.  The ``1370'' feature is a blend of \ion{O}{5}~$\lambda
1371$ and \ion{Fe}{5}~$\lambda\lambda 1360-1380$, while
\ion{Si}{3}~$\lambda 1417$, \ion{C}{3}~$\lambda 1427$, and
\ion{Fe}{5}~$\lambda 1430$ make up the ``1425'' feature.  This portion
of the spectra also encompasses the \ion{Si}{4}~$\lambda\lambda 1393,
1402$ doublet, which has a broad stellar P-Cygni component formed in
the expanding atmospheres of late O and early B supergiants;
superposed on this broad component are narrower interstellar
absorption lines due to the ambient interstellar medium which lies in
front of the stars.

In Figure~\ref{fig:uvzoom} we have also reproduced synthetic spectra
generated with the Starburst99 code \citep{sb99,llh+01} for the
standard case of continuous star formation and Salpeter initial mass
function, using in turn the two empirical stellar libraries available
in the code, built from spectra of OB stars in, respectively, the
Milky Way (the solar metallicity, or $Z = Z_{\odot}$, library) and the
Magellanic Clouds (the MC metallicity, or $Z = 1/3\, Z_{\odot}$,
library).  Figure~\ref{fig:uvzoom} shows that the former is a
plausible match to the composite spectrum of the higher stellar mass
galaxies, while the latter is consistent with the composite UV
spectrum of the galaxies in the lower stellar mass bins. We now
discuss this comparison in more detail.

Focusing on the composite spectrum of the most massive galaxies (the
lower panel in Figure~\ref{fig:uvzoom}), we find that the overall
equivalent widths of the ``1370'' and ``1425'' photospheric features
are in agreement between empirical and synthetic spectra. The
comparison of the \ion{Si}{4}~$\lambda\lambda 1393, 1402$ P-Cygni
component is complicated by the blending with the interstellar lines
which, as is usually the case, are stronger (and blue-shifted) in
starburst galaxies than in the individual stars which make up the
stellar libraries.  Nevertheless, the broad component of the
\ion{Si}{4} feature has comparable optical depth to the solar
metallicity Starburst99 model.  Turning to the lower mass galaxies
(upper panel in Figure~\ref{fig:uvzoom}), we see that all three
spectral features are undetected in the lower stellar mass UV
composite.  The Magellanic Cloud metallicity model spectrum also shows
these lines to be much reduced in strength.

The S/N ratio of the data and the subtlety of the spectral features in
question limit the above comparison to qualitative statements, and we
do not consider a more quantitative approach (such as a $\chi^2$
analysis, for example) to be warranted in the present
circumstances. Nevertheless, with only the UV spectra at our disposal,
we would have concluded that the galaxies with a mean stellar mass
$\langle M_{\star} \rangle = 7 \times 10^{10}$ \msun\ have metallicity
$Z \sim Z_{\odot}$ (or $12 + \log {\rm (O/H)} \sim 8.6$), and that
those with $\langle M_{\star} \rangle = 5 \times 10^{9}$ \msun\ have
$Z \lesssim 1/3\, Z_{\odot}$ (or $12 + \log {\rm (O/H)} \lesssim
8.1$).  Given the uncertainties, these conclusions are broadly
consistent with those deduced from our analysis of the
[\ion{N}{2}]/H$\alpha$ ratios.

There are other differences between the two composite UV spectra
reproduced in Figure 4; for example, the interstellar absorption lines
are significantly stronger in the galaxies with higher stellar mass.
The strongest UV interstellar lines, \ion{Si}{2}~$\lambda 1260$,
\ion{O}{1}~$\lambda 1302$+\ion{Si}{2}~$\lambda 1304$,
\ion{C}{2}~$\lambda 1334$, \ion{Si}{2}~$\lambda 1526$,
\ion{Fe}{2}~$\lambda 1608$, and \ion{Al}{2}~$\lambda 1670$, have
rest-frame equivalent widths $W_0 = 2 - 3$\,\AA\ in the $\langle
M^{\ast} \rangle = 7 \times 10^{10}$ \msun\ composite, and $W_0 = 1.5
- 2$\,\AA\ in the $\langle M^{\ast} \rangle = 5 \times 10^{9}$
\msun\ composite.  Since these lines are all strongly saturated
(e.g.\ \citealt{prs+02}), the higher equivalent widths are much more
likely to be due to a larger velocity dispersion of the absorbing gas,
than to an increase in the column densities of the metals.  This may
be related to differences in the star-formation histories of the two
samples of galaxies; on average, star formation has been in progress
for longer in galaxies with higher values of $\langle M_{\star}
\rangle$ and presumably more kinetic energy has been deposited in
their interstellar media, stirring the gas to higher velocity
dispersions.

Finally, the most obvious difference between the two
composite UV spectra is in the strengths of the two nebular emission
lines which fall within our wavelength range,
Ly$\alpha$ and C~{\sc iii}]~$\lambda 1908$, which are much stronger in
the lower stellar mass galaxies.  
Qualitatively, this difference is in agreement with the metallicity
and kinematics differences discussed above.
The C~{\sc iii}]~$\lambda \lambda 1907, 1909$ doublet is similar to
[O~{\sc iii}]~$\lambda\lambda 5007, 4959$
in showing a steep \emph{inverse} dependence on metallicity in the
high metallicity regime---as the metallicity increases, the temperature
of the H~{\sc ii} regions decreases and so do the relative populations
of the collisionally excited levels from which these emission lines
originate (e.g. \citealt{pp04}).
Thus, it is not surprising to find that
C~{\sc iii}]~$\lambda 1908$ is strong in the UV composite with
metallicity $Z \lesssim 1/3 Z_{\odot}$, and below our detection
limit in the galaxies with metallicities close to solar.
The dominant factor which determines the escape of
resonantly scattered Ly$\alpha$ photons from a star-forming galaxy
is the velocity dispersion of the interstellar gas through which the
photons propagate---the larger the velocity dispersion, the smaller
(and more blue-shifted) is the fraction of Ly$\alpha$ photons that
escape before being absorbed by dust and turned into infrared
photons (e.g. \citealt{mkt+03}).
Thus, the stronger Ly$\alpha$ flux of the
galaxies making up the lower stellar mass composite
in the top panel of Figure 4 goes hand in hand
with the smaller velocity dispersion of their interstellar
media, as discussed above.

\begin{figure}[htbp]
\plotone{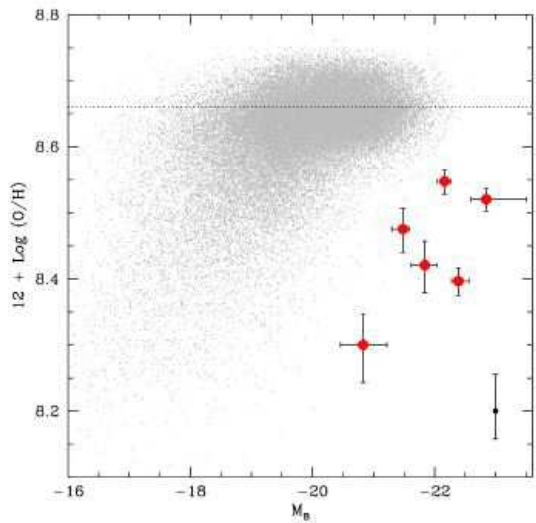}
\caption{The luminosity-metallicity relation at $z\gtrsim2$.  We have
  divided the sample into six bins by rest-frame absolute $B$
  magnitude, and estimated the metallicity in each bin.  The
  symbols are the same as in Figure~\ref{fig:metalmass}.  The points
  are not significantly correlated, though the faintest galaxies do
  have the lowest metallicities.  The lack of correlation can be
  understood through the large variation in the optical mass-to-light
  ratio at high redshift.  The \ztwo\ galaxies are $\sim3$
  magnitudes brighter than the SDSS galaxies at a given metallicity, as
  estimated by the \Ntwo/\Ha\ ratio.}
\label{fig:lumz}
\end{figure}

\begin{figure}[htbp]
\plotone{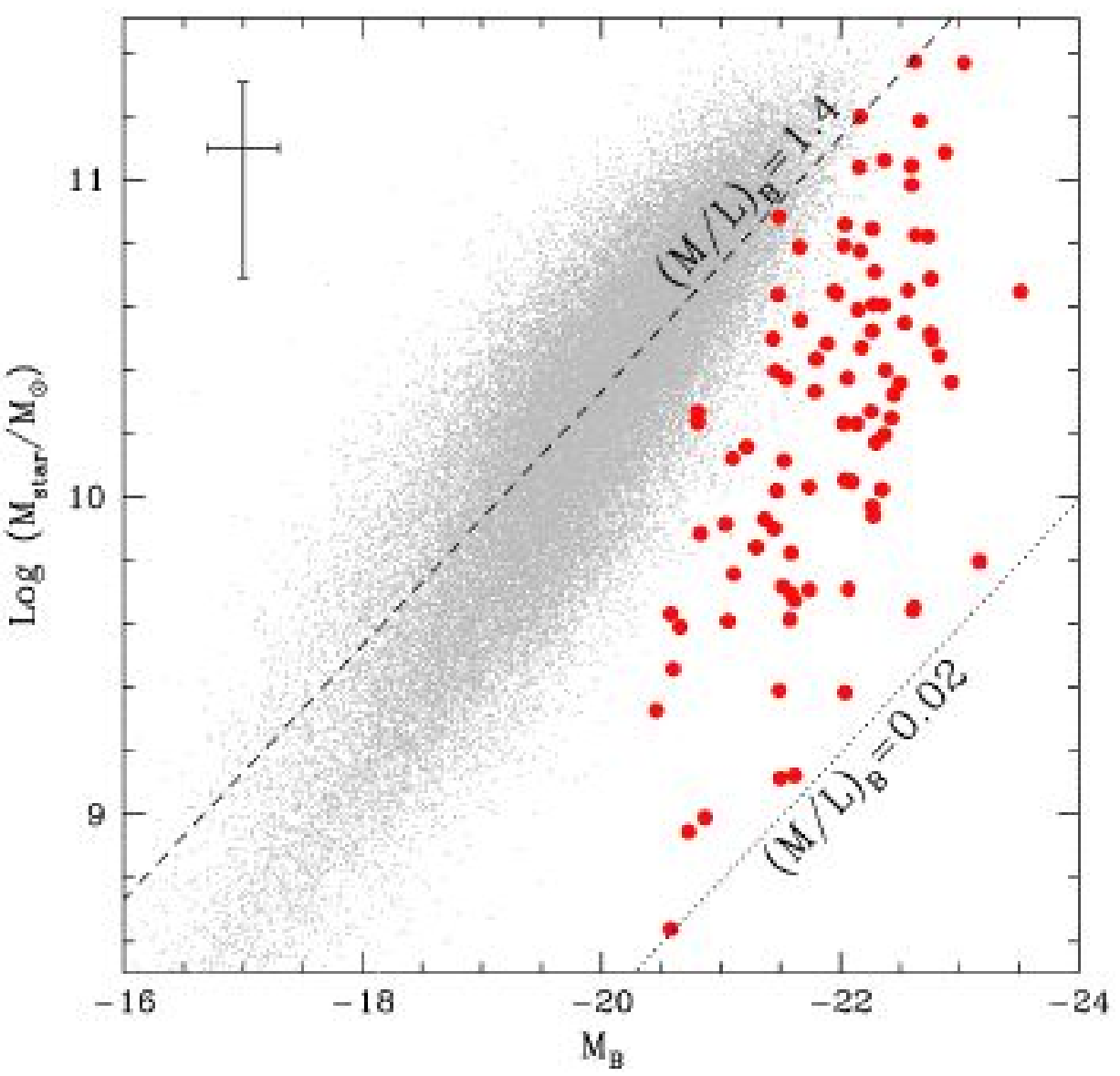}
\caption{Absolute rest-frame $B$ magnitude vs.\ stellar mass, for the
  individual galaxies in the \ztwo\ sample (large red circles) and the
  SDSS (small grey points).  The dashed and dotted lines show constant
  mass-to-light ratios $M/L$; in the \ztwo\ sample, the total range in
  luminosity is narrow and $M/L$ varies by a factor of $\sim70$ at
  most luminosities, while for most SDSS galaxies, $M/L$ varies by a
  factor of 2--5 at a given luminosity.}
\label{mlratio}
\end{figure}

\subsection{The Luminosity-Metallicity Relation}
Because the luminosity of a galaxy is far more easily determined than
its mass, there are a great many more luminosity-metallicity (L-Z)
relations than mass-metallicity relations in the literature
(e.g.\ \citealt{skh89,zkh94,gss97,lmc+04,slm+05}, to name only a few).
These correlations span 11 orders of magnitude in luminosity and 2 dex
in metallicity, and are seen in galaxies of all types.  Both the slope
and the zeropoint of the relation shift depending on the bandpass in
which the correlation is determined \citep{slm+05}; it is traditional
to use the absolute $B$ magnitude, but both the slope and the
dispersion of the relation decrease as wavelength increases to the IR,
probably due to decreased extinction and the closer correspondence of
the infrared luminosity to stellar mass.  The metallicity-luminosity
relation has been observed in galaxies at redshifts up to $z\sim1$
\citep{kwp+03,lcs03,kk04,mmh04,lhf+04}, at which point the shifting of
most of the strong nebular lines into the IR makes spectroscopy much
more difficult.  Most of these studies show that the zeropoint of the
L-Z relation evolves with redshift, so that galaxies of a given
luminosity have decreasing metallicity with increasing redshift.  The
so-far small number of metallicity-luminosity comparisons at $z>2$
confirm this trend, as high redshift galaxies are 2--4 mags brighter
than local galaxies of comparable metallicity
\citep{kk00,pss+01,sep+04}.

We construct a luminosity-metallicity relation analogous to our
mass-metallicity relation by dividing our sample
of 87 galaxies into six bins by rest-frame absolute $B$ magnitude
$M_B$, which we determine by multiplying the best-fit SED of each
object by the redshifted $B$-band transmission curve.  We construct a
composite spectrum of the galaxies in each bin, and measure the
\Ntwo/\Ha\ ratio and determine the oxygen abundance in the manner
described in \S\ref{sec:metalmeas}.  The results are shown in
Figure~\ref{fig:lumz}, again including SDSS metallicities determined from
the $N2$ indicator.  It is immediately apparent that the correlation
between luminosity and metallicity is weaker than that between mass
and metallicity; the trend with luminosity is not monotonic, and is
not statistically significant.  A Spearman correlation test finds a 33\%
probability that the points are uncorrelated, giving a significance of
1$\sigma$.  The faintest galaxies do have the
lowest metallicities, however; any appearance of correlation is driven by
this bin.

The comparison with the SDSS sample is again made difficult by the
saturation of the \Ntwo/\Ha\ ratio around solar metallicity, but it is
clear that the high redshift galaxies have both lower metallicities
(subject to the caveats discussed in \S\ref{sec:massmet}) and higher
luminosities than most of the local sample.  Considering the offset of
the more reliable lower metallicity bins, we see from
Figure~\ref{fig:lumz} that the \ztwo\ galaxies are approximately three
magnitudes brighter than local galaxies with the same oxygen
abundance.  It is difficult to invert the comparison to determine the
difference in metallicity between galaxies of a given luminosity in
the two samples, however, since virtually all of the local galaxies as
bright as the \ztwo\ sample have solar or greater abundances which
cannot be accurately determined by the $N2$ method.  In order for the
best-fit L-Z relation determined by T04 to pass through the average
luminosity and metallicity of our sample, it must be shifted downward
by $\sim0.9$ dex.  After allowing for the $\sim0.25$ dex systematic
difference between the metallicity diagnostics used by T04 and here
(as discussed in \S\ref{sec:massmet}), we are still left with a shift
of 0.6--0.7 dex between the local and high redshift samples.  Our mean
values are offset from other local L-Z relations by amounts ranging
from $\sim0.4$ \citep{skh89} to $\sim1.0$ \citep{lmc+04} dex.  The
large offsets between the various local relations are due to
differences in calibration methods and sample selection; the relations
are not directly comparable, and it is not yet clear which of them, if
any, provides the most appropriate comparison for our sample.  While
it is probably a robust conclusion that the \ztwo\ galaxies are more
metal-poor than their luminous local counterparts, a better
understanding of systematic effects is required to reliably quantify
the differences.

Our results are consistent with previous L-Z determinations at $z>2$,
and, moreover, they are not surprising given our knowledge of the high
redshift sample.  Star-forming galaxies at $z\sim2$ have lower average
mass-to-light ratios $M/L$ than local galaxies, and the variation in
$M/L$ at a given rest-frame optical luminosity can be as much as a
factor of $\sim70$ \citep{sse+05}.  This large variation in $M/L$
explains the lack of correlation in the \ztwo\ L-Z relation compared
to the local relation.  Figure~\ref{mlratio} shows absolute magnitude
$M_B$ plotted against stellar mass for the individual galaxies in the
\ztwo\ and SDSS samples. The \ztwo\ sample is shown by large red
points; at a given luminosity, $M/L$ varies by up to a factor of 70,
and the range in luminosity is small, so that galaxies with a wide
range of stellar masses fall in each of the six bins in luminosity.
In contrast, $M_{\star}$ and $M_B$ are much more tightly correlated in
the SDSS sample (small grey points); at a given luminosity, most of
the points are within a factor of a few in $M/L$.  If the fundamental
correlation is between metallicity and mass, these differences in
$M/L$ would shift the relation to higher luminosities and dramatically
increase the scatter, as we have observed.  A mass-metallicity
correlation is clearly more physically meaningful than a
luminosity-metallicity correlation at high redshifts; a corollary is
that the local metallicity-luminosity relation is simply a result of
the strong correlation between mass and luminosity at low redshift.
While it is likely that an L-Z relation determined in the rest-frame
IR would show a stronger correlation, investigation of this
possibility must wait until a larger sample of high $z$ galaxies with
both mid-IR photometry and nebular line spectra has been assembled.

\section{The Origin of the Mass-Metallicity Relation}
\label{sec:disc}
A correlation between gas-phase metallicity and stellar mass can
plausibly be explained either by the tendency of lower mass galaxies
to have larger gas fractions \citep{md97,bd00} and thus be less
enriched, or by the preferential loss of metals from galaxies with
shallow potential wells by galactic-scale winds.  With the relevant
information on the star, gas, and metal content of the galaxies, the
two effects can be differentiated.  In the simple, closed-box model of
chemical evolution with no inflows or outflows, the mass fraction of
metals $Z$ is a simple function of the gas fraction $\mu \equiv M_{\rm
  gas}/(M_{\rm gas}+M_{\star})$ and the true yield $y$, which
represents the ratio of the mass of metals produced and ejected by
star formation to the mass of metals locked in long-lived stars and
remnants.  The true yield is a function of stellar nucleosynthesis,
and as in previous, similar studies (T04, \citealt{g02}) we assume
that it is constant (see \citealt{g02} for a discussion).  The
metallicity is then given by
\begin{equation}
 Z=y\;{\rm ln}(1/\mu).  
\label{eq:yield}
\end{equation}
This equation can be inverted to determine the effective yield $y_{\rm
  eff}$ from the observed metallicity and gas fraction, $y_{\rm
  eff}=Z/ {\rm ln}(1/\mu)$.  If the simple model applies, $y_{\rm
  eff}$ will be constant for all masses and equal to the true yield
$y$, while a decrease in $y_{\rm eff}$ (either with respect to the
expected true yield or, more commonly, as a function of mass) is a
signature of outflows or of dilution by the infall of metal-poor gas
\citep{e90}.

T04 determined the effective yields of the SDSS galaxies by
using the empirical Schmidt law \citep{k98schmidt}, which relates the star
formation rate per unit area to the gas surface density, to estimate gas
masses.  They found lower effective yields in galaxies
with lower baryonic masses (the baryonic mass is expected to correlate
with the dark matter content, and thus indicate the depth of the
potential well; \citealt{msb+00}), and interpreted this result as
evidence for the preferential loss of 
metals in low-mass galaxies.  We carry out a similar analysis on our
sample of $z\sim2$ galaxies.  As described by \citet{ess+06a}, we
use a galaxy's \Ha\ luminosity and the spatial extent of its
\Ha\ emission $r_{\Ha}$ (after deconvolution with the seeing point
spread function) to calculate the \Ha\ luminosity per unit area
$\Sigma_{\Ha}$.  We then estimate each galaxy's gas surface density 
$\Sigma_{\rm gas}$ by combining the \citet{k98schmidt} relation
between star formation rate per unit area and gas density with the
conversion from \Ha\ luminosity to SFR from the same paper as follows:

\begin{equation}
\Sigma_{\rm gas}=1.6\times10^{-27}\left( \frac{\Sigma_{\rm
    H\alpha}}{{\rm erg\; s^{-1}\; kpc}^{-2}}\right)^{0.71} \; {\rm
  M_{\odot}\; pc^{-2}}.
\end{equation}
The estimated gas mass is then $M_{\rm gas}\sim\Sigma_{\rm
  gas}r_{\Ha}^2$.  We next combine $M_{\rm gas}$ and $M_{\star}$ to
obtain an estimate of the gas fraction $\mu$, and compute the mean
value of $\mu$ in each mass bin.  Although the considerable (factor of
$\sim2$) uncertainties in both the corrected \Ha\ flux and the
galaxy's size translate into a significant uncertainty in $\mu$ for
individual objects, the number of galaxies in each bin allows us to
determine the mean value of $\mu$ to within $\sim10$\%.

Galaxies at low and intermediate redshifts show significant
correlations between metallicity and extinction
(e.g.\ \citealt{cbb+05,mlc+05}).  Galaxies at \ztwo\ also show a
strong correlation between dust obscuration and stellar mass (and
hence metallicity), as shown by \citet{rsf+06}, who use the 24
\micron\ luminosity as observed by the Spitzer Space Telescope to
infer the infrared luminosity $L_{\rm IR}$ and the extinction as
parameterized by $L_{\rm IR}/L_{\rm UV}$.  One implication of the work
of \citet{rsf+06} is that $E(B-V)$ as determined from the UV slope may
overestimate the extinction correction for galaxies with ages less
than 100 Myr.  This effect can be seen in the high mean value of
$E(B-V)$ for the lowest mass bin in Table~\ref{tab:props}; for
galaxies with ages greater than 100 Myr in the current sample,
$E(B-V)$ and stellar mass are correlated with 4$\sigma$ significance.
This overestimation of the extinction in young galaxies has a
negligible effect on our current analysis; we estimate that it will
lead to an overestimation of the SFRs by a typical factor of
$\sim1.2$, considerably less than other uncertainties.  Only the
lowest mass bin is affected, since it is the only one that contains a
significant number of young objects; because we derive only an upper
limit on the mean metallicity of the galaxies in this bin, it does not
affect our conclusions.

As in the local universe \citep{md97,bd00}, there is a strong trend of
decreasing gas fraction with increasing stellar mass.  The lowest mass
bin in the sample has a mean gas fraction $\langle \mu \rangle =
0.85$, while the highest mass bin has $\langle \mu \rangle = 0.22$.
The median value of $\mu$ in our sample is $\sim0.5$, significantly
higher than the corresponding median value of $\sim0.2$ for the SDSS
galaxies considered by T04.  The low metallicities, low stellar masses
and high gas fractions of the objects in the lowest mass bin suggest
that they are young objects just beginning to form stars; this is also
indicated by the ages from the SED modeling (see
Table~\ref{tab:props}).  The correlations between age and other
properties are discussed in detail by \citet{ess+06a} in the context
of the comparison of stellar and dynamical masses.

\begin{figure}[htbp]
\plotone{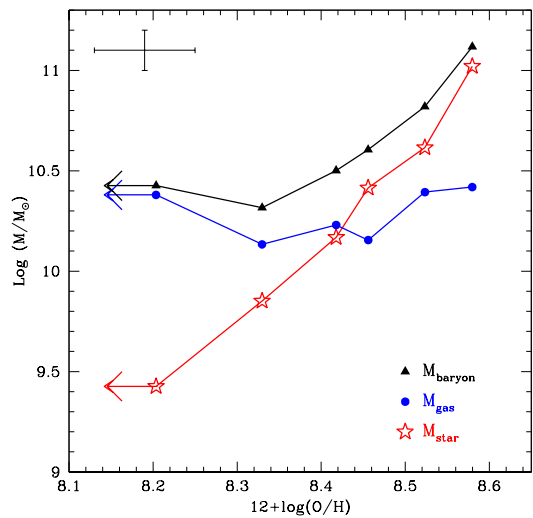}
\caption{The variation of stellar, gas, and baryonic mass in each of
  the six bins with metallicity.  Across the observed range
  in oxygen
  abundance, stellar mass increases strongly, baryonic
mass increases weakly, and gas mass remains approximately constant.
We thus see an increase in
metallicity with decreasing gas fraction.  The strong correlation
between stellar mass and age means that metallicity also increases
with the age of the stellar population.  The error bars at upper left
show typical uncertainties in gas and baryonic masses and
metallicities; uncertainties in stellar masses are smaller.}
\label{fig:massfrac}
\end{figure}

A consequence of the large gas fractions of the low stellar mass bins
is that the baryonic mass $M_{\rm gas}+M_{\star}$ of the galaxies in
the sample spans a much smaller range than the stellar mass.  The
difference in mean stellar mass between the highest and lowest mass
bins is a factor of 39, while the difference in mean baryonic mass
between the same bins is only a factor of six.  This relatively small
range in baryonic mass also limits our ability to detect differential
loss of metals as a function of mass.  We show the variations of
$M_{\star}$, $M_{\rm gas}$, and $M_{\rm bar}$ with metallicity in
Figure~\ref{fig:massfrac}.  Notably, the increase in baryonic mass
along our sequence of six bins is driven almost entirely by the
increase in stellar mass, while the gas mass remains relatively
constant.  This is because the average star formation rate and
\Ha\ luminosity vary much less than the average stellar mass across
the bins; in particular, galaxies in the lowest mass bin have higher
than average SFRs.  This may seem contrary to the trend of increasing
SFR at brighter $K$ magnitudes found by \citet{res+05}, but we account
for the difference both because we are less likely to detect
\Ha\ emission in $K$-faint galaxies unless it is especially strong (as
discussed by \citealt{ess+06a,ess+06b}) and because galaxies are more
likely to be detected in the $K$-band if they have high SFRs.
\citet{rsf+06} also show that the correlation of stellar mass and
bolometric luminosity is relatively weak, and that low stellar mass
galaxies span a wide range in $L_{\rm bol}$.  For present purposes,
there is a lower limit on the gas masses we are able to detect
corresponding to our \Ha\ flux limit, and the absence of galaxies with
both low stellar masses and low gas masses is probably a selection
effect.  We also note that the dynamical masses derived from the
\Ha\ line widths are a better match to the baryonic masses than the
stellar masses, with $M_{\rm dyn} \gg M_{\star}$ for the objects in
the lowest mass bin.  \citet{ess+06a} discuss this comparison in
detail.

\begin{figure*}[htbp]
\plotone{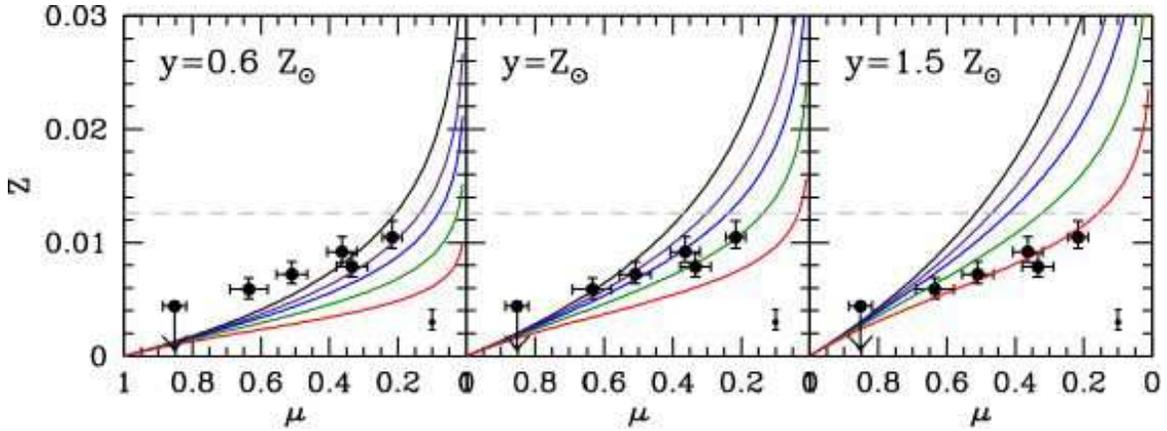}
\caption{The points in each panel show the mean metallicity $Z$ in
  each mass bin, plotted against the mean gas fraction $\mu$.  Gas
  fraction decreases from left to right, to show the increase in
  metallicity with decreasing gas fraction.  The curves illustrate the
  variation of $Z$ with $\mu$ for three different values of the true
  yield: from left to right, $y=0.6\,Z_{\odot}$ (our observed
  effective yield),  $y=Z_{\odot}$, and
  $y=1.5\,Z_{\odot}$.  Within each panel, the
  curves show varying values of the mass outflow rate $\dot M$,
  parameterized as a fraction $f$ of the star formation rate.  From
  top to bottom in each panel, the curves correspond to $f=0$ (black;
  the closed box model),
  0.5 (purple), 1 (blue), 2 (green) and 4 (red).  The data are best
  matched by a model with supersolar yield and an outflow rate $\dot
  M\sim4\times$SFR.  The horizontal dashed grey line corresponds to
  solar metallicity, and the small error bar in the lower right corner
  of each panel shows the systematic uncertainty in the metallicity
  calibration.} 
\label{fig:zevolve2}
\end{figure*}

The mean gas fractions and effective yields for each mass bin are
given in Table~\ref{tab:metals}.  In contrast to the SDSS sample,
there is no decline in the effective yield with decreasing baryonic
mass; in fact we see the opposite, as $y_{\rm eff}$ increases slightly
with decreasing mass.  The points in each panel of
Figure~\ref{fig:zevolve2} show the metallicity $Z$ in each mass bin
plotted against the mean gas fraction $\mu$, with decreasing gas
fraction from left to right to show the increase in $Z$ as $\mu$
declines.  We first discuss the data in the context of the closed box
model with no inflows or outflows.  For such a model, lines of
constant yield are curves given by Eq.~\ref{eq:yield}.  The uppermost
(black) curve on each of the three plots shows the variation of $Z$
with gas fraction in this model for three different values of the true
yield $y$; from left to right we show $y$ equal to our observed
effective yield $y_{\rm eff}=0.008\sim0.6\,Z_{\odot}$, solar yield
$y=0.0126$, and a supersolar yield $y=1.5\, Z_{\odot}$.  The points
are formally consistent with the $y\sim0.6\,Z_{\odot}$ model in the
left panel, as the measured value of $y_{\rm eff}$ in each bin is
within 1$\sigma$ of the weighted mean $y_{\rm eff}= 0.008$ in 4 of the
bins, and within 2$\sigma$ of this value in the fifth (we do not
consider the lowest mass bin, for which we find only an upper limit on
$Z$ and $y_{\rm eff}$); however, the systematic tendency of the lower
mass points to fall above the black line indicates that the model is
not a good fit. A $\chi^2$ test confirms this, giving a value of 9.

Given the ubiquitous signature of galactic-scale outflows in the
kinematics of the \ztwo\ galaxies, we now consider the effects of such
outflows on the metallicities of the galaxies.  To address this
question we modify the simple model to include gas outflow at rate
$\dot M$ \msunyr, which is a fraction $f$ of the star formation rate;
note that in this model the outflow rate does not directly depend on
the mass of the galaxy.  We assume that the metallicity of the
outflowing gas is the same as the metallicity of the gas that remains
in the \ion{H}{2} regions in the galaxy.  It can then be shown that
the metallicity is given by
\begin{equation}
Z=y\;(1+f)^{-1}\; {\rm ln}\;[1+(1+f)(\mu^{-1}-1)].
\label{eq:zoutflow}
\end{equation} 
Observations at low and high redshift suggest that $\dot M$ is
comparable to the star formation rate, and may be higher
(e.g.\ \citealt{m99,psa+00,m03}).  The curves in each panel of
Figure~\ref{fig:zevolve2} show the evolution of $Z$ with gas fraction
for, from top to bottom, $f=0$ (black; the closed box model), 0.5
(purple), 1 (blue), 2 (green), and 4 (red). A decrease in $y_{\rm
  eff}$ with decreasing gas fraction such as we have inferred is a
general feature of models with constant $\dot M/$SFR; more
specifically, from comparison with the points in each panel, it is
apparent that the data are best matched by a model with supersolar
yield $y\sim1.5\,Z_{\odot}$ and a high outflow rate $\dot
M\sim4\times$SFR (the red line in the righthand panel).  We caution
that the uncertainties in both metallicity and gas fraction are too
large to distinguish between the above models with confidence;
nevertheless the best-fitting model is not implausible, as we discuss
below.

An outflow rate of several times the star formation rate has already
been suggested by observations of the the lensed LBG MS1512-cB58;
\citet{prs+02} found a value of $\dot M$ higher than the star
formation rate using conservative assumptions for the size of the
galaxy and outflow velocity.  This scenario is also in good agreement
with observations and predictions of the enrichment of the
intracluster medium (ICM).  The ICM contains several times more mass
in baryons than is found in the cluster galaxies themselves, and thus
it contains most of the metals in clusters as well
(e.g.\ \citealt{ml97}).  Recent models for the enrichment of the ICM
\citep{dkw04,nlb+05} use feedback from supernovae to transport metals
out of galaxies, finding that enrichment of the ICM occurs at high
redshift, with of order half of the metals produced by \ztwo, and that
half or more of the metals in the ICM are produced in massive galaxies
($L\gtrsim L_{\ast}$, \citealt{nlb+05}; or baryonic mass $\gtrsim
10^{10}\, h^{-1}$ \msun, \citealt{dkw04}).  These models also require
either a top-heavy IMF in starbursts or a significantly enhanced yield
in order to reproduce the observed metal abundances in clusters; some
such variant IMF is a feature of many models of star formation and
chemical enrichment in clusters
(e.g.\ \citealt{zs96,fbb03,tbm+04}). The supersolar yield suggested by
the best-fitting model shown in Figure~\ref{fig:zevolve2} may support
such a scenario, although we emphasize that our data do not require an
IMF with more high-mass stars than the \citet{c03} IMF we use here.
Such an IMF, when combined with estimates of yields from stellar
nucleosynthesis (e.g.\ \citealt{mm02}), results in a metal yield that
is solar or higher.

\subsection{Redshift Evolution of the Mass-Metallicity Relation}
\label{sec:evol}
With estimates of the gas fractions in hand we now return to the
$\sim0.3$ dex offset between the local and \ztwo\ mass-metallicity
relations.  There are several possible reasons for this factor of
$\sim2$ difference.  The magnitude of the offset itself is not
well-determined because of the uncertainty in our metallicity scale
indicated by the offsets between the \Othree/\Hb\ and
\Ntwo/\Ha\ ratios relative to the SDSS galaxies seen in
Figure~\ref{fig:diag}.  The effects of these offsets cannot be
quantified reliably without detailed modeling of a larger sample of
galaxies with all four emission lines, but meanwhile we can use
several techniques to estimate the possible uncertainty in metallicity
that may result.  We have another method of determining the
metallicity of the four galaxies with all four emission lines in the
$O3N2$ index of \citet{pp04}: $O3N2\equiv \log$ \{[\ion{O}{3}]$\lambda
5007/\rm H\beta) / ($[\ion{N}{2}$]\lambda 6583/\rm H\alpha)$\}.
\citet{pp04} find that $O3N2$ varies with oxygen abundance as
\begin{equation}
{\rm 12 + log\; (O/H)} = 8.73 - 0.32 \times O3N2.
\end {equation}
Using this calibration to determine the metallicities of the four
galaxies with all four lines, we find oxygen abundances that are 0.17
dex lower on average than those determined using \Ntwo/\Ha.  This is
one plausible estimate of the metallicity uncertainty due to the
differing physical conditions in the \ion{H}{2} regions of the
\ztwo\ galaxies.  We can obtain another by assuming that the observed
shift in the \Othree/\Hb\ and \Ntwo/\Ha\ diagram is entirely in the
\Ntwo/\Ha\ direction; then the relatively shallow slope of the
relationship between \Ntwo/\Ha\ and (O/H) leads to a maximum offset in
metallicity of $\sim0.3$ dex.  The effect of an offset in the
\Othree/\Hb\ and \Ntwo/\Ha\ diagram on metallicity determinations is
considered in detail by \citet{scmb05}, and we refer the reader to
Section 5.3 of that paper for a discussion.  These authors consider
the effect of a harder ionizing spectrum in high redshift galaxies,
and conclude that this could plausibly lead to a factor of
$\sim1.5$--2 uncertainty in metallicity.

Another possible issue is that the SDSS metallicities are biased
toward the innermost regions of the galaxies observed, where the
spectrograph fibers were positioned.  According to a recent
re-analysis of this effect by \citet{ek05}, this fact alone can
account for a $\sim0.15$ dex offset between nuclear and global (i.e.,
integrated over the whole galaxy) metallicities.  However, this effect
may be mitigated by the fact that our global spectra are also biased
toward the galaxies' central regions, which have the highest surface
brightness.  We conclude that the uncertainty in the metallicity
offset between the \ztwo\ and local galaxies is approximately a factor
of two, about the same as the offset itself.

There are reasons to believe that there may be a real evolutionary
offset in the mass-metallicity relation, and we proceed with the
discussion under this assumption.  As described above, we find that
the galaxies in our sample have an average gas fraction more than two
times higher than the galaxies studied by T04, placing them at an
earlier stage in the process of converting their gas to stars.  Given
this less evolved state, it is not surprising that they should have
lower gas-phase metallicities.  Redshift evolution of the
mass-metallicity relation has also recently been investigated by
\citet{sgl+05} with a sample of galaxies at $0.4<z<1$, finding that a
galaxy of a given stellar mass tends to have lower metallicity at
$z\sim0.7$ than at $z\sim0$.  Using the SDSS relation of T04, their
sample at $z\sim0.7$, and the \ztwo\ galaxies presented by
\citet{sep+04}, they develop an empirical model for the redshift
evolution of the relation.  This model predicts metallicities an
average of $\sim0.2$ dex higher than we observe, though at least some
of this offset is likely to be due to the different metallicity
indicators used (it is not possible to place our sample and that of
\citet{sgl+05} on the same metallicity scale, as an accurate
conversion between the two indicators used has not yet been
established for the samples in question).  Such a shift to lower
metallicities (or higher masses) with increasing redshift can
qualitatively be explained via a scenario in which star formation
takes place over a more protracted period in lower mass galaxies.
This scenario is consistent with our results; we find that the more
massive galaxies in our sample have smaller gas fractions, and are
thus likely to exhaust their gas supply before the less massive
galaxies.

The \citet{sgl+05} model also predicts a steeper relation at
\ztwo\ than we have observed; this is a consequence of their
assumption that the shape of the T04 relation remains unchanged but
shifts to higher masses, whereas we find that our data is
well-approximated by the T04 relation shifted to lower metallicities
at the same stellar mass.  In physical terms, the slope of the stellar
mass-metallicity relation depends on both the yield and the presence
or absence of outflows (and inflows).  As shown in
Figure~\ref{fig:zevolve2}, a closed box model such as that assumed by
\citet{sgl+05} results in a steep relation between metallicity and gas
fraction (and stellar mass), while the \ztwo\ data are better
described by the shallower slope of a model with significant outflows.

While comparisons between the metallicities of star-forming galaxies
at high and low redshift are of obvious interest, it is very likely
that our sample and that of T04 do not form an evolutionary sequence.
The likely descendants of the \ztwo\ sample can be identified by
comparing their clustering properties, evolved to $z\sim0$, with those
of objects in the local universe; such a comparison shows that
early-type galaxies in the SDSS match the clustering properties of the
\ztwo\ sample, while the later-type star-forming objects studied by
T04 are too weakly clustered to be the descendants of the
\ztwo\ population \citep{asp+05}.  In this sense it is more relevant
to compare the metallicities of the \ztwo\ galaxies with local
early-type objects than with star-forming galaxies at $z\sim0$, though
such a comparison would undoubtedly be complicated by systematic
offsets resulting from the very different methods used to determine
metal abundances in elliptical galaxies (usually absorption features
in the integrated spectra of old stellar populations).  Broadly
speaking, however, most early-type galaxies in the SDSS with
$M_{\star}>10^{10}$ \msun\ have metallicities ranging from $Z\sim0.6\,
Z_{\odot}$ to $Z\sim1.5\, Z_{\odot}$ \citep{gsb+05}; these results are
consistent with the probable final stellar masses and metallicities of
the \ztwo\ galaxies.

Substantial uncertainties in the mass-metallicity relation at
\ztwo\ remain, of course.  It is important to confirm the trend in
metallicity with additional measurements and with abundance indicators
that use a broader set of emission lines.  The absolute values of the
abundances in the sample are uncertain, and the improved understanding
of the galaxies' physical conditions that will result from a larger
sample of lines will be essential for determining this absolute scale.
Furthermore, our derivation of the gas masses and gas fractions is
indirect, and assumes that the Schmidt law takes the same form at
\ztwo\ as in the local universe.  This has not yet been tested,
although the one similar galaxy with a direct measurement of the gas
mass, the lensed $z=2.7$ LBG MS1512-cB58, appears to be consistent
with the local Schmidt law \citep{btg+04}.  As long as some form of
the Schmidt law applies, our results will be qualitatively similar.  A
related question concerns the appropriateness of the gas masses
derived from the Schmidt law.  These represent only the gas associated
with current star formation, and are therefore almost certainly an
underestimate of the total gas masses.  In a typical disk galaxy
today, $\sim40$\% of the gas mass is not included by the Schmidt law
\citep{mk01}; this fraction could plausibly be higher in the young
starbursts in our sample.  It is not clear how or if this gas affects
the metal enrichment and star formation.  Given our lack of
information on this gas, we do not consider it.  T04 discussed these
questions of gas masses derived from the Schmidt law in somewhat more
detail, but arrived at similar conclusions.

Our simple model for the effect of winds on gas-phase abundances
assumes that the metallicity of the outflowing gas is the same as the
observed abundances in the galaxy; this may not be correct.
Metal-enhanced hot winds have been observed in X-rays in local
starbursts; in dwarf galaxies, such winds may carry away nearly all
the oxygen produced by the burst \citep{mkh02}.  Such a metal-enhanced
outflow would increase metal losses for a given outflow rate; thus it
may be possible to produce the relative shallow relationship between
$Z$ and $\mu$ that we have observed with a mass outflow rate smaller
than the value $\dot M\sim4\times$SFR we find above.

We also consider the possibility, discussed in \S\ref{sec:massmeas}
and \S\ref{sec:massmet}, that we have underestimated the stellar
masses of the galaxies with the smallest $M_{\star}$ by a factor of a
few.  This would primarily affect the objects in the lowest mass bins,
and would result in a decrease in the gas fraction and a decrease in
the effective yield.  A factor of three increase in the stellar mass
of the galaxies in the two lowest mass bins would result in a decrease
in $y_{\rm eff}$ of approximately a factor of two for those bins; this
is enough to remove the observed trend of decreasing $y_{\rm eff}$
with increasing stellar mass, and results instead in an approximately
constant effective yield, roughly consistent with the closed box
model.  Given the significant evidence for strong outflows in these
galaxies, however, we regard the best-fit model obtained above as more
plausible.

Finally, the sample of \ztwo\ galaxies considered here is by no means
complete.  The UV-selection technique is most likely to miss galaxies
that are very dusty or have little current star formation;
\citet{res+05} show that there is relatively little overlap
($\sim12$\%) between the UV-selected sample and \ztwo\ galaxies
selected by their $J-K_s>2.3$ colors (Distant Red Galaxies, or DRGs;
\citealt{flr+03}).  This technique selects galaxies with a strong
optical break, due to either an evolved stellar population or strong
reddening (e.g.\ \citealt{pmd+05}).  Little is known about the
metallicities of the DRGs.  A few rest-frame optical spectra of
brighter DRGs with $K<20$ have presented by \citet{vff+04,vkr+05};
these suggest approximately solar metallicities, although they
frequently show evidence for AGN or shock ionization.  It will be
extremely difficult to measure metallicities for passively evolving
galaxies at high redshift, since they lack strong emission lines.
Massive galaxies that have consumed or expelled most of their gas
might be expected to be among the most metal-rich objects at high
redshift.  Dusty, rapidly star-forming galaxies are likely to have
large gas masses, but their metallicities will depend on how many
generations of stars have enriched the gas.  There is probably a range
of metallicities within this population.  Some confirmation of this
can be found in Figure 15 of \citet{rsf+06}, who plot inferred gas
fraction against stellar mass for near-IR selected galaxies as well as
for UV-selected galaxies similar to those considered here.  All
samples considered follow the same trend of decreasing gas fraction
with increasing stellar mass, with the DRGs at the high mass and low
gas fraction end of the spectrum.  This suggests that galaxies
selected by near-IR techniques may also follow a similar
mass-metallicity relation to that discussed here, with a possible
extension to higher metallicities for the oldest and most massive
objects.  It will be interesting to see if this proves to be true when
larger samples of metallicity measurements become available.

In summary, the \ztwo\ galaxies show a strong trend in oxygen
abundance over a range of only a factor of $\sim6$ in baryonic mass.
The effective yield increases slightly with decreasing mass, rather
than decreasing as would be expected if low mass galaxies lost a
larger fraction of their metals to outflows.  We conclude that the
mass-metallicity relation at high redshift is primarily a product of
varying stages of galaxy evolution, caused by the increase in
metallicity as gas is converted into stars and metals are returned to
the remaining gas.  It may be modulated by metal loss from strong
outflows in galaxies of all masses, which results in a shallower
increase of metallicity with decreasing gas fraction than predicted by
the closed box model.  There is no evidence for preferential loss of
metals from lower mass galaxies, as has been inferred in the local
universe, although the small mass range spanned by our sample limits
our sensitivity to such an effect.

\section{Summary and Conclusions}
\label{sec:end}

We have used composite \Ha\ + \Ntwo\ spectra of 87 star-forming galaxies at
$z\gtrsim2$, divided into six bins by stellar mass, to study the
correlation between stellar mass and metallicity at high redshift.
Our conclusions are summarized as follows.

1.  There is a strong correlation between stellar mass and metallicity
at $z\gtrsim2$, as the quantity $12+\rm log(O/H)$ increases monotonically from
$<8.2$ for galaxies with $\langle M_{\star} \rangle = 2.7 \times
10^{9}$ \msun\ to 8.6 for galaxies with $\langle M_{\star} \rangle = 1.0 \times
10^{11}$ \msun.   The relation is offset by $\sim0.3$ dex from the
local mass-metallicity relation, in the sense that galaxies of a given
stellar mass have lower metallicities at high redshift.  The absolute
values of the oxygen abundances are uncertain, but the trend
is unlikely to be due to systematic effects 
such as AGN contamination, hidden stellar mass, or variations in the
ionization parameter of the \ion{H}{2} regions.

2.  Rest-frame $B$-band luminosity and metallicity are not
significantly correlated.  The \ztwo\ galaxies are systematically
brighter than local star-forming galaxies spanning the same range in
stellar mass, showing that they have smaller mass-to-light ratios
$M/L$.  At a given metallicity, local galaxies are $\sim3$ magnitudes
fainter than the \ztwo\ sample.  The known large scatter in the
rest-frame optical $M/L$ at \ztwo\ accounts for the lack of
correlation between luminosity and metallicity, and indicates that the
correlation with stellar mass is (as expected) more fundamental.

3.  We use the Schmidt law to estimate the gas masses and gas
fractions of the \ztwo\ galaxies, finding that the gas fraction
increases substantially with decreasing stellar mass.  The lowest mass
bin in the sample has a mean gas fraction of 85\%, while the highest
stellar mass bin has a fraction of 20\%.  Our median gas fraction is
$\sim50$\%, as compared to $\sim20$\% in local star-forming galaxies.
Galaxies with low stellar masses, low metallicities and high gas
fractions also have young ages.  A consequence of the trend in gas
fraction with stellar mass is a much smaller range in baryonic mass (a
factor of 6) than stellar mass (a factor of 39) across the sample.

4.  The observed metallicities and gas fractions allow an estimate of
the effective yield in each mass bin.  In contrast to the results of a
similar study in the local universe, we find no decrease in $y_{\rm
  eff}$ at low masses; instead, $y_{\rm eff}$ increases slightly with
decreasing mass.  Qualitatively, such an increase is a feature of
models in which galaxies of all masses lose metals from outflows.
More quantitatively, comparison with simple models shows that the
observed variation of metallicity with gas fraction is best described
by a model with supersolar yield and an outflow rate $\sim4$ times
higher than the star formation rate.  We conclude that the
mass-metallicity relation at high redshift is caused by the increase
in metallicity as gas is converted to stars, and may be modulated by
strong outflows in galaxies of all masses.  Our ability to detect
differential metal loss as a function of mass is limited by the small
range of baryonic masses spanned by the galaxies in the sample, but
there is no evidence for preferential loss of metals from low mass
galaxies as is suggested locally.

Much remains to be done in order to improve upon the substantial
uncertainties inherent in the present work.  Independent measurements
of the gas masses of galaxies at high redshift, though very difficult,
are essential to determine whether our derived gas fractions and
effective yields are valid.  Additional metallicity measurements,
based on other indicators that use a wider set of emission lines, are
needed to confirm the trend with stellar mass revealed by the $N2$
index, to provide a better understanding of the physical conditions in
the \ion{H}{2} regions, and to establish a secure absolute calibration
of the metallicity scale.  Further investigation of the question of
differential metal loss from outflows at high redshift requires
observations of fainter galaxies with smaller potential wells, to
expand the dynamic range in baryonic mass.  We anticipate that all of
these measurements will greatly increase our understanding of the
interplay between stars and gas, within and outside of galaxies, at
high redshift.

\acknowledgements We thank the anonymous referee for useful comments
and a careful reading of the paper. We also thank Mike Dopita, Lisa
Kewley and Evan Skillman for illuminating discussions, and the staffs
of the Keck and Palomar observatories for their assistance with the
observations. CCS, DKE and NAR have been supported by grant
AST03-07263 from the US National Science Foundation and by the David
and Lucile Packard Foundation.  AES acknowledges support from the
Miller Institute for Basic Research in Science, and KLA from the
Carnegie Institution of Washington.  Finally, we wish to extend
special thanks to those of Hawaiian ancestry on whose sacred mountain
we are privileged to be guests.  Without their generous hospitality,
most of the observations presented herein would not have been
possible.


\clearpage

\begin{landscape}
\input{tab1.tex}
\input{tab2.tex}
\clearpage
\end{landscape}

\end{document}

%% file: tab1.tex
\begin{deluxetable}{l l l l l l l l l l}
\tablewidth{0pt}
\tabletypesize{\footnotesize}
\tablecaption{Mean Stellar Masses and Stellar Population Properties\label{tab:props}}
\tablehead{
\colhead{Bin} & 
\colhead{N\tablenotemark{a}} & 
\colhead{$\langle z \rangle$\tablenotemark{b}} & 
\colhead{${\cal R}$\tablenotemark{c}} &
\colhead{$K_s$\tablenotemark{d}} &
\colhead{$M_B$\tablenotemark{e}} &
\colhead{Stellar Mass\tablenotemark{f}} & 
\colhead{SFR\tablenotemark{g}} & 
\colhead{Age\tablenotemark{h}} & 
\colhead{E(B-V)\tablenotemark{i}} \\
\colhead{} &
\colhead{} &
\colhead{} &
\colhead{} &
\colhead{} &
\colhead{} &
\colhead{($10^{10}$ \msun)} &
\colhead{(\msunyr)} &
\colhead{(Myr)} &
\colhead{}
}
\startdata
1 & 15 & 2.36 & 24.20 $\pm$ 0.72 & 21.51 $\pm$ 0.69 & -21.27 $\pm$ 0.73 & 0.27 $\pm$ 0.15 & 55 $\pm$ 52 & 180 $\pm$ 276 & 0.19 $\pm$ 0.09 \\
2 & 15 & 2.27 & 23.96 $\pm$ 0.81 & 21.03 $\pm$ 0.61 & -21.66 $\pm$ 0.59 & 0.71 $\pm$ 0.17 & 24 $\pm$ 10 & 448 $\pm$ 374 & 0.14 $\pm$ 0.07\\
3 & 15 & 2.21 & 24.01 $\pm$ 0.72 & 20.86 $\pm$ 0.47 & -21.82 $\pm$ 0.58 & 1.5 $\pm$ 0.3 & 30 $\pm$ 21 & 968 $\pm$ 944 & 0.11 $\pm$ 0.08\\
4 & 14 & 2.28 & 23.97 $\pm$ 0.62 & 20.54 $\pm$ 0.49 & -22.15 $\pm$ 0.51 & 2.6 $\pm$ 0.4 & 28 $\pm$ 17 & 1026 $\pm$ 864 & 0.16 $\pm$ 0.07\\
5 & 14 & 2.21 & 23.93 $\pm$ 0.60 & 20.16 $\pm$ 0.38 & -22.33 $\pm$ 0.51 & 4.1 $\pm$ 0.6 & 47 $\pm$ 30 & 1311 $\pm$ 790 & 0.17 $\pm$ 0.09\\
6 & 14 & 2.26 & 24.32 $\pm$ 0.55 & 20.05 $\pm$ 0.40 & -22.30 $\pm$ 0.44 & 10.5 $\pm$ 5.4 & 47 $\pm$ 24 & 2409 $\pm$ 591 & 0.19 $\pm$ 0.06

\enddata
\tablenotetext{a}{Number of galaxies in bin}
\tablenotetext{b}{Mean and standard deviation of $z_{\Ha}$.}
\tablenotetext{c}{Mean and standard deviation of ${\cal R}$ magnitude.}
\tablenotetext{d}{Mean and standard deviation of $K_s$ magnitude.}
\tablenotetext{e}{Mean and standard deviation of absolute magnitude
  $M_B$, determined as described in the text.}
\tablenotetext{f}{Mean and standard deviation of stellar mass from SED fitting; we use
a \citet{c03} IMF.}
\tablenotetext{g}{Mean and standard deviation of SFR from extinction-corrected \Ha\ luminosity,
  including a factor of two aperture correction determined from
  narrow-band imaging and comparison of the $K$-band continuum with
  broad-band magnitudes.  Because the dispersion is large we also quote the semi
  interquartile range for bins 1--6 respectively: 29, 7, 13, 12, 13
  and 15 \msunyr. }
\tablenotetext{h}{Mean and standard deviation of best-fit age from SED
  fitting.  Because the dispersion is large we also quote the semi
  interquartile range for bins 1--6 respectively: 123, 184, 677, 490,
  541 and 375 Myr.}
\tablenotetext{i}{Mean and standard deviation of best-fit $E(B-V)$ from SED
  fitting.}

\end{deluxetable}

%% file: tab2.tex
\begin{deluxetable}{l c c c c c c c c}
\tablewidth{0pt}
\tabletypesize{\footnotesize}
\tablecaption{Oxygen Abundances and Gas Fractions\label{tab:metals}}
\tablehead{
\colhead{Bin} & 
\colhead{Stellar Mass\tablenotemark{a}} & 
\colhead{$F_{\Ha}$\tablenotemark{b}} &
\colhead{$F_{[\rm NII]}$\tablenotemark{b}} &
\colhead{$N2$\tablenotemark{c}} &
\colhead{12 + log(O/H)\tablenotemark{d}} & 
\colhead{$M_{\rm bar}$\tablenotemark{e}} & 
\colhead{$\mu_{\rm gas}$\tablenotemark{f}} &
\colhead{$y_{\rm eff}$\tablenotemark{g}}\\
\colhead{} &
\colhead{($10^{10}$ \msun)} &
\colhead{($10^{-17}$ erg s$^{-1}$ cm$^{-2}$)} &
\colhead{($10^{-17}$ erg s$^{-1}$ cm$^{-2}$)} &
\colhead{} &
\colhead{} &
\colhead{($10^{10}$ \msun)} &
\colhead{} &
\colhead{} 
}
\startdata
1 & 0.27 $\pm$ 0.15 & $20.5\pm0.5$ & $<1.2$ & $<-1.22$ &  $<8.20$ & 2.7$\pm$1.7 & 0.85 $\pm$ 0.12 & $<0.027$ \\
2 & 0.71 $\pm$ 0.17 & $13.9\pm0.3$ & $1.4\pm0.2$ & $-1.00^{+0.07}_{-0.09}$ & $8.33^{+0.07}_{-0.07}$ & 2.1$\pm$0.6 & 0.63 $\pm$ 0.12 & 0.013$\pm$0.003 \\
3 & 1.5 $\pm$ 0.3 & $18.7\pm0.4$ & $2.7\pm0.3$ & $-0.85^{+0.05}_{-0.06}$ & $8.42^{+0.06}_{-0.05}$ & 3.2$\pm$1.1 & 0.48 $\pm$ 0.19 & 0.010$\pm$0.002\\
4 & 2.6 $\pm$ 0.4 & $15.9\pm0.4$ & $2.6\pm0.3$ & $-0.78^{+0.05}_{-0.05}$ & $8.46^{+0.06}_{-0.05}$ & 4.0$\pm$0.9 & 0.33 $\pm$ 0.12 & 0.007$\pm$0.001\\
5 & 4.1 $\pm$ 0.6 & $24.3\pm0.5$ & $5.3\pm0.4$ & $-0.66^{+0.03}_{-0.04}$ & $8.52^{+0.06}_{-0.05}$ & 6.6$\pm$1.1 & 0.36 $\pm$ 0.10 & 0.009$\pm$0.002\\
6 & 10.5 $\pm$ 5.4 & $27.0\pm0.4$ & $7.4\pm0.3$ & $-0.56^{+0.02}_{-0.02}$ & $8.58^{+0.06}_{-0.04}$ & 13.1$\pm$5.6 & 0.22 $\pm$ 0.11 & 0.007$\pm$0.001
\enddata
\tablenotetext{a}{Mean and standard deviation of stellar mass from SED fitting; we use
a \citet{c03} IMF.}
\tablenotetext{b}{Fluxes of \Ha\ and \Ntwo$\lambda 6584$ from the composite
  spectra.}
\tablenotetext{c}{$N2\equiv {\rm log} (F_{[\rm NII]}/F_{\Ha})$}
\tablenotetext{d}{Oxygen abundance from $N2$, using the
  calibration of \citet{pp04}.}
\tablenotetext{e}{Mean and standard devation of the baryonic mass $M_{\rm gas} + M_{\star}$, with gas masses determined from the Schmidt law as described in the text.} 
\tablenotetext{f}{Mean and standard deviation of the gas fraction $\mu
  = M_{\rm gas}/(M_{\rm gas} + M_{\star}$).}
\tablenotetext{g}{Effective yield $y_{\rm eff}=Z/{\rm ln}(1/\mu)$.}
\end{deluxetable}